\documentclass[preprint]{imsart}
\pdfoutput=1
\usepackage{amsthm,amsmath,natbib}

\startlocaldefs
\usepackage[margin=1.5in]{geometry}

\usepackage{amssymb}
\usepackage{graphicx}
\usepackage{pdflscape}
\usepackage{makecell}
\usepackage{listings}
\usepackage{float}
\usepackage{courier}
\usepackage{hyperref}
\usepackage{booktabs}
\usepackage{rotating}
\usepackage{algorithm2e}
\usepackage{chngcntr}
\usepackage{afterpage}
\usepackage{tikz}
\usepackage{xr-hyper}
\makeatletter
\newcommand*{\addFileDependency}[1]{
  \typeout{(#1)}
  \@addtofilelist{#1}
  \IfFileExists{#1}{}{\typeout{No file #1.}}
}
\makeatother

\usepackage[above]{placeins}
\allowdisplaybreaks
\usepackage{caption}
\usepackage{subcaption}
 
\usepackage{color}
\usepackage{fancyvrb}

\DefineVerbatimEnvironment{Highlighting}{Verbatim}{commandchars=\\\{\}}
\usepackage{framed}
\definecolor{shadecolor}{RGB}{248,248,248}
\newenvironment{Shaded}{\begin{snugshade}}{\end{snugshade}}

\newcommand{\DecValTok}[1]{\textcolor[rgb]{0.00,0.00,0.81}{{#1}}}

\newcommand{\StringTok}[1]{\textcolor[rgb]{0.31,0.60,0.02}{{#1}}}

\newcommand{\ImportTok}[1]{{#1}}
\newcommand{\CommentTok}[1]{\textcolor[rgb]{0.56,0.35,0.01}{\textit{{#1}}}}

\newcommand{\FunctionTok}[1]{\textcolor[rgb]{0.00,0.00,0.00}{{#1}}}
\newcommand{\VariableTok}[1]{\textcolor[rgb]{0.00,0.00,0.00}{{#1}}}

\newcommand{\OperatorTok}[1]{\textcolor[rgb]{0.81,0.36,0.00}{\textbf{{#1}}}}
\newcommand{\BuiltInTok}[1]{{#1}}
\newcommand{\ExtensionTok}[1]{{#1}}

\newcommand{\NormalTok}[1]{{#1}}
 
\newfloat{lstfloat}{htbp}{ext}
\floatname{lstfloat}{Listing}

\setcitestyle{nameyear}

\usepackage{letltxmacro}
\LetLtxMacro\oldttfamily\ttfamily
\DeclareRobustCommand{\ttfamily}{\oldttfamily\csname ttsize\endcsname}

\newcommand{\pluseq}{\mathrel{+}=} 

\usepackage[para]{threeparttable}

\makeatletter
\def\TPT@doparanotes{\par
   \prevdepth\z@ \TPT@hsize
   \TPTnoteSettings
   \parindent\z@ \pretolerance 8
   \linepenalty 200
   \renewcommand\item[1][]{\relax\ifhmode \begingroup
       \unskip
       \advance\hsize 10em 
       \penalty -45 \hskip\z@\@plus\hsize \penalty-19
       \hskip .15\hsize \penalty 9999 \hskip-.15\hsize
       \hskip .01\hsize\@plus-\hsize\@minus.01\hsize 
       \hskip 0em\@plus .3em
      \endgroup\fi
      \tnote{##1}\,\ignorespaces}%
   \let\TPToverlap\relax
   \def\endtablenotes{\par}%
}

\makeatother

\endlocaldefs

\begin{document}

\begin{frontmatter}

\title{High-Performance Statistical Computing in the Computing Environments of the  2020s}
\runtitle{High-performance statistical computing}


\author{\fnms{Seyoon} \snm{Ko}\ead[label=e1]{kos@ucla.edu}%
\thanksref{t1}}
\thankstext{t1}{This article is partly based on the first author's doctoral  dissertation \citep{Ko2020thesis}.} 
\author{\fnms{Hua} \snm{Zhou}}\ead[label=e2]{huazhou@ucla.edu}
\affiliation{Department of Biostatistics, UCLA Fielding School of Public Health, Los Angeles, CA, USA}
\author{\fnms{Jin J.} \snm{Zhou}}\ead[label=e3]{jinjinzhou@ucla.edu}
\affiliation{Department of Medicine, UCLA David Geffen School of Medicine, Los Angeles, CA, USA and\\
Department of Epidemiology and Biostatistics, College of Public Health, University of Arizona, Tucson, AZ, USA}
\author{\fnms{Joong-Ho} \snm{Won}\corref{}\ead[label=e4]{won.j@snu.ac.kr}}
\affiliation{Department of Statistics, Seoul National University, Seoul, Korea}

\address{Department of Biostatistics, University of California, Los Angeles, e-mail:\texttt{\emph{kos@ucla.edu; huazhou@ucla.edu}}.}
\address{Departments of Medicine, University of California, Los Angeles, e-mail:\texttt{\emph{jinjinzhou@ucla.edu}}.}
\address{Departments of Statistics, Seoul National University, e-mail:\texttt{\emph{won.j@snu.ac.kr}}.}

\runauthor{S. Ko, H. Zhou, J. J. Zhou, and J.-H. Won}
\begin{abstract}
Technological advances in the past decade, hardware and software alike, have made access to high-performance computing (HPC) easier than ever. We review these advances from a statistical computing perspective. Cloud computing makes access to supercomputers affordable. Deep learning software libraries make programming statistical  algorithms easy and enable users to write code once and run it anywhere --- from a laptop to a workstation with multiple graphics processing units (GPUs) or a  supercomputer in a cloud. 
Highlighting how these developments benefit statisticians, 
we review recent optimization algorithms that are
useful for high-dimensional models and can harness the power of HPC.
Code snippets are provided 
to demonstrate the ease of programming.
We also provide an easy-to-use distributed matrix data structure suitable for HPC. Employing this data structure, we illustrate various  statistical applications including large-scale positron emission tomography and $\ell_1$-regularized Cox regression. Our examples easily scale up to an 8-GPU workstation and a 720-CPU-core cluster in a cloud. As a case in point, we analyze the onset of type-2 diabetes  from the UK Biobank with 200,000 subjects and about 500,000 single nucleotide polymorphisms using the HPC  $\ell_1$-regularized Cox regression. Fitting this half-million-variate model takes less than 45 minutes and reconfirms known associations. 
To our knowledge, this 
is the first demonstration of the feasibility of penalized regression of survival outcomes at this scale.
\end{abstract}
\begin{keyword}
\kwd{high-performance statistical computing}
\kwd{graphics processing units (GPUs)}
\kwd{cloud computing}
\kwd{deep learning}
\kwd{MM algorithms}
\kwd{ADMM}
\kwd{PDHG}
\kwd{Cox regression}
\end{keyword}
\end{frontmatter}
\newpage
\section{Introduction}\label{sec:intro}
Clock speeds of the central processing units (CPUs) on the desktop and laptop computers hit the physical limit more than a decade ago, and there will likely be no major breakthrough until quantum computing becomes practical. Instead, the increase in computing power is now accomplished by using multiple 
cores within a processor chip.
High-performance computing (HPC) means computations that are so large that their requirement on storage, main memory, and raw computational speed cannot be met by a single (desktop) computer \citep{hager2010introduction}.
Modern HPC machines are equipped with more than one CPU that can work on the same problem \citep{eijkhout2013introduction}. 
Often, special-purpose co-processors such as graphics processing units (GPUs) are attached to the CPU 
to improve the speed by orders of magnitude for certain tasks.
First developed for rendering graphics on a computer screen, 
a GPU can be thought of a massively parallel matrix-vector multiplier and vector transformer on a data stream.
%
With increasing needs to analyze petabyte-scale data, the success of large-scale statistical computing relies on efficiently engaging HPC in the statistical practice. 

About a decade ago, the second author discussed the potential of GPUs in statistical computing: 
\citet{zhou2010graphics} predicted that ``GPUs will fundamentally alter the landscape of computational statistics.'' Yet, it does not appear that GPU computing, or HPC in general, has completely permeated the statistical community. Part of the reason for this may be attributed to the fear that parallel and distributed code is difficult to program, especially in R \citep{rcoreteam}, the \textit{lingua franca} of statisticians.%
\footnote{Although there exist several R packages for high-performance computing \citep{cranhpcctv}, their functionalities and 
usability appear not to match what is available in other languages. In particular, the authors were not able to come up with a simple implementation of the computational tasks presented in this paper without writing low-level C/C++ code or using an interface to Python.}
On the other hand, the landscape of scientific computing in general, including so-called data science \citep{donoho2017}, has indeed substantially changed. Many high-level programming languages, 
such as Python \citep{rossum1995python} and Julia \citep{bezanson2017julia}, support parallel computing by design or through standard libraries. Accordingly, many software tools have been developed in order to ease programming in and managing HPC environments. Last but not least, cloud computing \citep{fox2011cloud} is getting rid of the necessity for purchasing expensive supercomputers and scales computation as needed.

Concurrently, easily parallelizable algorithms for fitting statistical models with hundreds of thousand parameters have also seen significant advances. 
Traditional Newton-Raphson or quasi-Newton type of algorithms face two major challenges in contemporary problems: 1) explosion of dimensionality renders storage and inversion of Hessian matrices prohibitive; 2) regularization of model complexity is almost essential in high-dimensional settings, which is often realized by nondifferentiable penalties; this leads to high-dimensional, nonsmooth optimization problems. 
For these reasons, nonsmooth first-order methods have been extensively studied during the past decade \citep{beck2017first},
since Hessian matrix inversion can be completely avoided.
For relatively simple, decomposable penalties \citep{negahban2012unified}, the proximal gradient method \citep{Beck:SiamJournalOnImagingSciences:2009,combettes2011proximal,parikh2014proximal,polson2015proximal} produces a family of easily parallelizable algorithms. For the prominent example of the Lasso \citep{tibshirani1996regression}, this method contrasts to the highly efficient sequential coordinate descent method of \citet{friedman2010regularization} and  smooth approximation approaches, e.g., \citet{hunter2005variable}. Decomposability or separability of variables is often the key to parallel and distributed algorithms. 
The 
alternating direction method of multipliers 
\citep[ADMM,][]{gabay1976dual,Boyd:FoundationsAndTrendsInMachineLearning:2010}
achieves this goal through variable splitting, while often resulting in nontrivial subproblems to solve. 
As an alternative, the primal-dual hybrid gradient (PDHG) algorithm  \citep{zhu2008efficient,Esser:SiamJournalOnImagingSciences:2010,chambolle2011first,Condat:JournalOfOptimizationTheoryAndApplications:2012,Vu2013} has a very low per-iteration complexity, useful for complex penalties such as the generalized lasso \citep{Tibshirani:TheAnnalsOfStatistics:2011,ko2019easily,ko2019optimal}. 
Another route 
toward separability is 
the 
majorization-minimization (MM)
principle \citep{lange2000optimization,hunter2004tutorial,lange2016mm}, which has been explored in \citet{zhou2010graphics}. In fact, the proximal gradient method can be viewed as a realization of the MM principle. Recent developments in the application of this principle include 
distance majorization \citep{chi2014distance} and proximal distance algorithms \citep{keys2016proximal}. 
When the matrix to be inverted to solve the optimality condition has many independent components, nonsmooth Newton methods \citep{kummer1988newton,qi1993nonsmooth} can be a viable option; see \citet{jin2019unified} for recent applications to sparse regression. Nonsmooth Newton methods can also be combined with first-order methods for more complex nonsmooth penalties \citep{chu2020semismooth,won2020proximity}.%

The goal of this paper is to review the advances in parallel and distributed computing environments during the past decade and demonstrate how easy it has become to write code for large-scale, high-dimensional statistical models and run it on various distributed environments. In order to make the contrast clear, we deliberately take examples from \citet{zhou2010graphics}, namely positron emission tomography (PET), nonnegative matrix factorization (NMF), and multidimensional scaling (MDS). 
The difference lies in the scale of the examples: our experiments deal with data of size at least $10,000 \times 10,000$ and as large as $200,000 \times 200,000$ for dense data, and $810,000\times 179,700$ for sparse data. This contrasts with the size of at best $4096 \times 2016$ of \citet{zhou2010graphics}. This level of scaling is possible because the use of \textit{multiple} GPUs in a distributed fashion has become handy, as opposed to the single GPU, 
C-oriented programming environment of 2010.
Furthermore, using the power of cloud computing and modern deep learning software, we show that exactly the \textit{same}, easy-to-write code can run on multiple CPU cores and/or clusters of workstations. Thus we bust the common misconception that deep learning software is dedicated to neural networks and heuristic model fitting. Wherever possible, we apply more recent algorithms in order to cope with the scale of the problems. 
In addition, a new example of large-scale proportional hazards regression model is  investigated. We demonstrate the potential of our approach through a single
multivariate Cox regression model regularized by the $\ell_1$ penalty on the UK Biobank genomics data (with 200,000 subjects), featuring time-to-onset of Type 2
Diabetes (T2D) as outcome and 500,000 genomic loci harboring single
nucleotide polymorphisms as covariates. To our knowledge, such a
large-scale joint genome-wide association analysis has not been
attempted. The reported Cox regression model retains a large
proportion of \textit{bona fide} genomic loci associated with T2D and recovers
many loci near genes involved in insulin resistance and
inflammation, which may have been missed in conventional univariate analysis
with moderate statistical significance values.

The rest of this article is organized as follows.
We review HPC systems and see
how they have become easy to use in Section \ref{sec:environ}. 
In Section \ref{sec:lib}, we review software libraries employing the ``write once, run everywhere'' principle  (especially deep learning software) and discuss how they can be employed for fitting high-dimensional statistical models on the HPC systems of Section \ref{sec:environ}. 
In Section \ref{sec:algo}, we review modern scalable optimization techniques well-suited to  HPC environments. 
We present how to distribute a large matrix over multiple devices in Section \ref{sec:distmat}, and numerical examples in Section \ref{sec:num}. 
The article is concluded in Section \ref{sec:discuss}. 

\section{Accessible  High-Performance Computing Systems}\label{sec:environ}

\subsection{Preliminaries}\label{sec:environ:prelim}

Since modern HPC relies on parallel computing, in this section we review several concepts from parallel computing literature at a level  minimally necessary for the subsequent discussions. 
Further details can be found in \citet{nakano2012parallel,eijkhout2013introduction}.

\paragraph{Data parallelism.}
While parallelism can appear at various levels such as \linebreak instruction-level and task-level, what is most relevant to statistical computing is data-level parallelism or data parallelism. If data can be 
subdivided into several 
pieces that can be processed independently of each other, then we say there is  data parallelism in the problem. 
Many operations such as scalar multiplication of a vector, matrix-vector multiplication, and summation of all elements in a vector can exploit data parallelism using parallel 
architectures, which will be discussed shortly. 

\paragraph{Memory models.}
In any computing system, processors (CPUs or GPUs) need to access data residing in the memory. While \emph{physical} computer memory uses complex hierarchies (L1, L2, and L3 caches; bus- and network-connected, etc.), systems employ abstraction to provide programmers an appearance of  transparent memory access. Such \emph{logical} memory models can be categorized into the shared memory model and the distributed memory model. 
In the shared memory model, all processors share the \emph{address space} of the system's memory even if it is physically distributed. 
For example, when two processors refer to a variable $x$, the variable is stored in the same memory address. Hence, if one processor alters the variable, then the other processor is affected by the modified value.
Modern CPUs that have several cores within a processor chip fall into this category.
On the other hand, in the distributed memory model, the system has memory both physically and logically distributed.
Processors have their own memory address spaces and cannot see each other's memory directly. 
If two processors refer to a variable $x$, then there are two separate memory locations, each of which belongs to each processor under the same name.
Hence the memory does appear distributed to programmers, and some explicit communication mechanism is required in order for processors to exchange data with each other. 
The advantage at the cost of this complication is scalability --- the number of processors that can work in a tightly coupled fashion is much greater in distributed memory systems (say 100,000) than shared memory systems (say four, as many recent laptops are equipped with a CPU chip with 4 cores). Hybrids of the two memory models are also possible. A typical computer \emph{cluster} consists of multiple \emph{nodes} interconnected in a variety of network topology. 
A node is a workstation that can run standalone, with its main memory shared by several processors installed on the motherboard. Hence within a node, it is a shared memory system, whereas across the nodes the cluster is a distributed memory system.

\paragraph{Parallel programming models.}
For shared-memory systems, programming models based on \emph{threads} are most popular. 
A thread is a stream of machine language instructions that can be created and run in parallel during the execution of a single program.
OpenMP is a widely used extension to the C and Fortran programming languages based on threads. It achieves data parallelism by letting the compiler know what part of the sequential program is parallelizable by creating multiple threads. 
Simply put, each processor core can run a thread operating on a different partition of the data.
In distributed-memory systems, parallelism is difficult to  achieve via a simple modification of sequential code. 
The programmer needs to coordinate communications between processors not sharing memory.
A de facto standard for such processor-to-processor communication is the message passing interface (MPI). 
MPI routines mainly consist of \emph{point-to-point communication calls} that send and receive data between two processors, and \emph{collective communication calls} that all processors in a group participate in. Typical collective communication calls include
\begin{itemize}
    \item Scatter: one processor has a data array, and each other processor receives a partition of the array;
    \item Gather: one processor collects data from all the processors to construct an array;
    \item Broadcast: one processor sends its data to all other devices;
    \item Reduce: gather data and produce a combined output on a root process based on an associative binary operator, such as sum or maximum of all the elements.
\end{itemize}
%
There are also all-gather and all-reduce, where the output is shared by all processors.
At a higher abstraction level, MapReduce \citep{dean2008mapreduce}, a functional programming model in which a ``map'' function transforms each datum into a key-value pair, and a ``reduce'' function aggregates the results, is a popular distributed data processing model. While basic implementations are provided in base R, both the map and reduce operations are easy to parallelize. Distributed implementations such as Hadoop \citep{hadoop} handle communications between nodes implicitly. This programming model is inherently one-pass and stateless, and iterations on Hadoop require frequent accesses to external storage (hard disks), hence slow. Apache Spark \citep{zaharia2010spark} is an implementation that substitutes external storage with memory caching, yet iterative algorithms are an order of magnitude slower than their MPI counterparts \citep{jha2014tale,reyes2015big,gittens2016matrix}.

%

\paragraph{Parallel architectures.}
To realize the above models, a computer architecture that allows simultaneous execution of multiple machine language instructions is needed. Single instruction, multiple data (SIMD) architecture has multiple processors that execute the same instruction on different parts of the data. The GPU falls into this category of architectures, as its massive number of cores can run a large number of threads sharing memory. Multiple instruction, multiple data (MIMD), or single program, multiple data (SPMD) architecture has multiple CPUs that execute independent parts of program instructions on their own data partition. Most computer clusters fall into this category. 

\subsection{Multiple CPU nodes: clusters, supercomputers, and clouds}

%
Computing on multiple nodes can be utilized in many different scales. 
For mid-sized data, one may build his/her own cluster with a few nodes. 
This requires determining the topology and purchasing all the required hardware, along with resources to maintain it. This is certainly not an expertise of virtually all statisticians. 
Another option may be using a well-maintained supercomputer in a nearby HPC center. 
A user can take advantage of the facility with up to hundreds of thousand cores. 
The computing jobs on these facilities are often controlled by a job scheduler, such as Sun Grid Engine \citep{sge}, Slurm \citep{slurm}, and Torque \citep{torque}.
However, access to supercomputers is almost always limited. 
Even when the user has access to them, he/she often has to wait in a very long queue until the requested computation job is started by the scheduler. 

In recent years, cloud computing,
which refers to both the applications delivered as services over the Internet, and the hardware and systems software in the data centers that provide these services \citep{Armbrust:2010},
has emerged as a third option.
Information technology giants such as Amazon, Microsoft, and Google lend their practically infinite computing resources 
to users on demand 
by wrapping the resources as ``virtual machines,'' which are charged 
per CPU hours and storage.
Users basically pay utility bills for their use of computing resources.
An important implication of this infrastructure to end-users is that the cost of using 1000 virtual machines for one hour is almost the same as using a single virtual machine for 1000 hours. Therefore a user can build his/her own virtual cluster ``on the fly,'' increasing  the size of the cluster as the size of the problem to solve grows.
A catch here is that a cluster does not necessarily possess the power of HPC as suggested in Section \ref{sec:environ:prelim}:
a requirement for high performance is that all the machines should run in tight lockstep when working on a problem \citep{fox2011cloud}.
%
%
However, early cloud services
were more focused on web applications 
that do not involve frequent data transmissions between computing instances,
and less optimized for 
HPC, yielding discouraging results \citep{evangelinos2008cloud,walker2008benchmarking}. 
%
For instance, ``serverless computing'' services such as AWS Lambda, Google Cloud Functions, and Azure Functions
allow users to run a function on a large amount of data, in much the same fashion as supplying it to \texttt{lapply()} in base R.
These services offer reasonable scalability on a simple map-reduce-type jobs such as image featurization, word count, and sorting. Nevertheless, their restrictions on resources (e.g., single core and 300 seconds of runtime in AWS Lambda) and the statelessness of the functional programming approach result in high latency for iterative algorithms, such as consensus ADMM (Aytekin and Johansson, 2019).

Eventually,
many improvements have been made at hardware and software 
levels to make 
HPC on clouds feasible.
At hardware level, cloud service providers now support CPU instances such as \texttt{c4}, \texttt{c5}, and \texttt{c5n} instances of Amazon Web Services (AWS), with up to 48 physical cores of higher clock speed of up to 3.4 GHz along with support for accelerated SIMD computation.
If network bandwidth is critical, the user may choose instances with faster networking (such as \texttt{c5n} instances in AWS), allowing up to 100 Gbps of network bandwidth. 
At the software level, these providers support tools that manage  resources efficiently for scientific computing applications, such as ParallelCluster \citep{parallelcluster} and ElastiCluster \citep{elasticluster}. These tools are designed to run programs in clouds in a similar manner to 
proprietary clusters through a job scheduler. 
In contrast to a physical cluster in an HPC center,
a virtual cluster on a cloud 
is exclusively created for the user; there is no need for waiting in a long queue.
Consequently, over 10 percent of all HPC jobs are running in clouds, and over 70 percent of HPC centers run some jobs in a cloud as of June 2019; the latter is up from just 13 percent in 2011 \citep{hyperion:2019}.

In short, cloud computing is now a cost-effective option for statisticians who demand high performance, without a steep learning curve.




\subsection{Multi-GPU node}

In some cases, HPC is achieved by installing multiple GPUs on a single 
node.
%
A key feature of GPUs is their ability to apply a mapping to a large array of floating-point numbers simultaneously. 
The mapping (called a \emph{kernel}) can be programmed by the user. 
This feature is enabled by integrating a massive number of simple compute cores in a single processor chip, forming a SIMD architecture.
While this architecture of GPUs was created for high-quality video games to generate a large number of pixels in a hard time limit, the programmability and high throughput soon gained attention from the scientific computing community. 
Matrix-vector multiplication and elementwise nonlinear transformation of a vector can be computed several orders of magnitude faster on GPU than on CPU.
Early applications of general-purpose GPU programming include physics simulations, signal processing, and geometric computing \citep{owens2007survey}. 
Technologically savvy statisticians demonstrated its potential in 
Bayesian simulation \citep{suchard2010some,suchard2010understanding} and 
high-dimensional optimization
\citep{zhou2010graphics,yu2015high}. 
Over time, 
the number of cores has increased from 240 (Nvidia GTX 285, early 2009) to 4608 (Nvidia Titan RTX, late 2018)
and more local memory 
--- separated from CPU's main memory ---
has been added (from 1GB of GTX 285 to 24GB for Titan RTX). 
GPUs could only use single-precision for their floating-point operations, but they now support double- and half-precisions. More sophisticated operations such as tensor multiplication are also supported. 
High-end GPUs are now being designed specifically for scientific computing purposes, sometimes with fault-tolerance features such as error correction. 

Major drawbacks of GPUs are smaller memory size, compared to CPU, and data transfer overhead between CPU and GPU. 
%
These limitations can be addressed by using multiple GPUs: 
recent GPUs can be installed on a single node 
and communicate with each other without the meddling of CPU;
this effectively increases the local memory of a collection of GPUs.%
\footnote{\citet{lee2017large} explored this possibility in image-based regression.}
It is relatively inexpensive to construct a node with 4--8 desktop GPUs compared to a cluster of CPU nodes with a similar computing power (if the main computing tasks are well suited for the SIMD model), and 
the gain is much larger than the cost.
A good example would be linear algebra operations that frequently occur in high-dimensional optimization.

Programming environments for GPU computing have been notoriously hostile to programmers for a long time.
The major 
hurdle is that a programmer needs to write two suits of code, the \emph{host} code that runs on a CPU and \emph{kernel} functions that run on  GPU cores. Data transfer between  CPU and GPU(s) also has to be taken care of.
Moreover, kernel functions need to be written in special extensions of C, C++, or Fortran, e.g., the Compute Unified Device Architecture \citep[CUDA,][]{cuda} or Open Computing Language  \citep[OpenCL,][]{munshi2009opencl}.
Combinations of these technical barriers 
prevented casual programmers, especially statisticians, from writing GPU code despite its computational gains. 
There were efforts to sugar-coat these hostile environments with a high-level language
such as R \citep{buckner2009gputools} or Python \citep{tieleman2010gnumpy,klockner2012pycuda,lam2015numba}, but these attempts struggled to garner large enough user base 
since
the functionalities were often limited and inherently hard to extend. 

Fortunately, GPU programming environments have been revolutionized since deep learning \citep{lecun2015deep} brought sensation to many machine learning applications. Deep learning is almost synonymous to deep neural networks, which refer to a repeated (``layered'') application of an affine transformation of the input followed by identical elementwise transformations through a nonlinear link function, or ``activation function.''
Fitting a deep learning model is almost always conducted via (approximate) minimization of the specified loss function through a clever application of the chain rule to the gradient descent method, called ``backpropagation'' \citep{rumelhart1988learning}. These computational features fit well to the SIMD architecture of GPUs, use of which dramatically reduces the training time of this highly overparameterized family of models with a huge amount of training data \citep{raina2009large}. 
Consequently, many efforts have been made to ease GPU programming for deep learning, resulting in easy-to-use software libraries. Since the sizes of neural networks get ever larger, more HPC capabilities, e.g., support for multiple GPUs and CPU clusters, have been developed. As we review in the next section, programming with those libraries gets rid of many hassles with GPUs, close to the level of conventional programming.


%
\section{Easy-to-use Software Libraries for HPC}\label{sec:lib}

\subsection{Deep learning libraries and HPC}
As of 
revising
this article (summer 2020), the two most popular deep learning software libraries are TensorFlow \citep{tensorflow} and PyTorch \citep{paszke2017pytorch}. 
There are two common features of these libraries. One is the computation graph that automates the evaluation of the loss function and its differentiation required for backpropagation.
The other feature, more relevant to statistical computing, is an efficient and user-friendly interface to linear algebra and convolution routines that work on both CPU and GPU in a unified fashion. A typical pattern of using these libraries is to specify the model and describe how to fit the model to the training data in a high-level scripting language (mostly Python). 
The system on which the model is fitted can be programmed.
If the target system is a CPU node, then the software can be configured to utilize the OpenBLAS \citep{openblas} or Intel Math Kernel Library \citep{mkl}, which are optimized implementations of the Basic Linear Algebra Library \citep[BLAS,][]{blas} for shared-memory systems.
If the target system is a workstation with a GPU, then the same script can employ a pair of host and kernel code that may make use of cuBLAS \citep{cublas}, a GPU version of BLAS, and cuSPARSE \citep{cusparse}, GPU-oriented sparse linear algebra routines.
%
A slight change in the option for device selection --- usually a line or two in the script --- can control whether to run the model on a CPU or GPU.
From the last paragraph of the previous section, we see  that this ``write once, run everywhere'' feature of deep learning libraries can make GPU programming easier for statistical computing as well.

TensorFlow is a successor of Theano \citep{theano}, one of the first libraries to support automatic differentiation based on computational graphs. Unlike Theano, which generates GPU code on the fly, TensorFlow is equipped with pre-compiled GPU code for a large class of pre-defined operations.
%
%
%
%
PyTorch inherits Torch \citep{collobert2011torch7}, an early machine learning library written in a functional programming language called Lua, and Caffe \citep{jia2014caffe}, a Python-based deep learning library. 
PyTorch (and Torch) can also manage GPU memory efficiently. As a result, it is known to be faster than other deep learning libraries \citep{bahrampour2015comparative}. 

Both libraries support multi-GPU and multi-node computing.%
\footnote{%
There are other deep learning software libraries with similar HPC supports: 
Apache MxNet \citep{chen2015mxnet} supports multi-node computation via Horovod; multi-GPU computing is also supported at the interface level.
Microsoft Cognitive Toolkit \citep[CNTK,][]{seide2016cntk} supports parallel stochastic gradient algorithms through MPI.%
}
In TensorFlow, multi-GPU computation is supported natively on a single node. If data are distributed in multiple GPUs and one needs data from the other, the GPUs communicate with each other implicitly and the user does not need to interfere. 
For multi-node communication,
it is recommended to use MPI through the library called Horovod \citep{sergeev2018horovod} for  tightly-coupled HPC environments. 
In PyTorch, both multi-GPU and multi-node computing are enabled by using the interface \texttt{torch.distributed}.
This interface defines MPI-style (but simplified) communication primitives (see 
Section \ref{sec:environ:prelim}). 
Implementations
include the \textit{bona fide} MPI, Nvidia Collective Communications Library (NCCL), and Gloo \citep{gloo}.
Recent MPI implementations can map
multi-GPU communication to the MPI standard as well as traditional multi-node 
communication. While NCCL is useful for a multi-GPU node, Gloo is useful with multiple CPU with Ethernet interconnect.

\subsection{Automatic differentiation}\label{sec:ad}
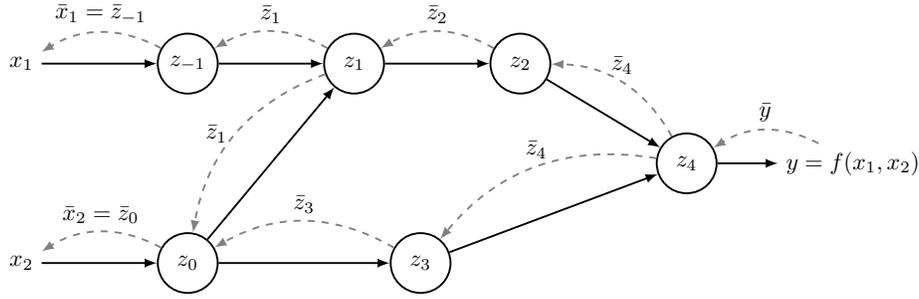
\begin{figure}
    \centering
    \resizebox{0.9\textwidth}{!}{
    \begin{tikzpicture}[]
	\normalsize
	\tikzstyle{vnode} = [circle,draw,thick,fill=white,minimum size=9mm]
	\tikzstyle{vedge} = [->,>=latex,thick]
	
	\node[vnode] (z-1) at (-8.5,0.5) {$z_{-1}$};
	\node[vnode] (z0) at (-8.5,-2.5) {$z_0$};
	\node[vnode] (z1) at (-6,0.5) {$z_1$};
	\node[vnode] (z3) at (-5.0,-2.5) {$z_3$};
	\node[vnode] (z2) at (-3.5,0.5) {$z_2$};
	\node[vnode] (z4) at (-1.0,-1.0) {$z_4$};
	
	\node[] (x1) at (-11,0.5) {$x_1$};
	\node[] (x2) at (-11,-2.5) {$x_2$};
	\node[] (f) at (1.5,-1) {$y=f(x_1, x_2)$};
	
	\draw (z-1) edge [vedge] (z1);
	\draw (z1) edge [bend right,vedge,gray,dashed] node[above,black] {$\bar{z}_1$} (z-1);
	\draw (z0) edge [vedge] (z1);
	\draw (z1) edge [bend right,vedge,gray,dashed] node[left,black] {$\bar{z}_1$} (z0);
	\draw (z0) edge [vedge] (z3);
	\draw (z3) edge [bend right,vedge,gray,dashed] node[above,black] {$\bar{z}_3$} (z0);
	\draw (z1) edge [vedge] (z2);
	\draw (z2) edge [bend right,vedge,gray,dashed] node[above,black] {$\bar{z}_2$} (z1);
	\draw (z3) edge [vedge] (z4);
	\draw (z4) edge [bend right,vedge,gray,dashed] node[above,black] {$\bar{z}_4$} (z3);
	\draw (z2) edge [vedge] (z4);
	\draw (z4) edge [bend right,vedge,gray,dashed] node[above,black] {$\bar{z}_4$} (z2);
	
	\draw (x1) edge [vedge] (z-1);
	\draw (z-1) edge [bend right,vedge,gray,dashed] node[above,black] {$\bar{x}_1 = \bar{z}_{-1}$} (x1);
	\draw (x2) edge [vedge] (z0);
	\draw (z0) edge [bend right,vedge,gray,dashed] node[above,black] {$\bar{x}_2 = \bar{z}_0$} (x2);
	\draw (z4) edge [vedge] (f);
	\draw (f) edge [bend right,vedge,gray,dashed] node[above,black] {$\bar{y}$} (z4);

\end{tikzpicture}
    }
    \caption{Computational graph for evaluating function $f(x_1, x_2) = \log(x_1 + x_2) - x_2^2$.
    Dashed arrows indicate the direction of backpropagation evaluating $\nabla f(x_1, x_2)$.%
    } 
    \label{fig:cgraph}
\end{figure}

The automatic differentiation (AD) feature of deep learning software deserves separate attention. AD refers to a collection of techniques that evaluate the derivatives of a function specified by a computer program accurately \citep{griewank2008evaluating,baydin2017automatic}.
Based on AD, complex deep models can be trained with stochastic approximation (see the next section) on huge data within a hundreds of lines of code
and approximate a rich class of functions efficiently;
see
\citet{schmidt2020,bauer2019,imaizumi2019,suzuki2019,ohn2019smooth} for recent theoretical developments.
Most AD techniques rely on decomposition of the target function into elementary functions (primitives) whose derivatives are known, and the computational graph, either explicitly or implicitly, that describes the dependency among the primitives. Fig. \ref{fig:cgraph} illustrates the computational graph for the bivariate function $f(x_1, x_2) = \log(x_1 + x_2) - x_2^2$. The internal nodes represent intermediate variables corresponding to the primitives: $z_{-1}=x_1$, $z_0=x_2$, $z_1 = z_{-1} + z_{0}$, $z_2=\log z_1$, $z_3=z_0^2$, and $z_4=z_2 - z_3$; $y=z_4$.

There are two modes of AD, depending on the order of applying the chain rule. Forward-mode AD applies the rule from right to left (or from input to output), hence it is straightforward to implement. In Fig. \ref{fig:cgraph}, if we want to evaluate the partial derivative $\frac{\partial f}{\partial x_2}$ at $(3,2)$, then by denoting $\dot{z}_i\equiv\frac{\partial z_i}{\partial x_2}$ we see that $\dot{z}_{-1}=\dot{x}_1=0$, $\dot{z}_0=\dot{x}_2=1$, $\dot{z}_1=\dot{z}_0 + \dot{z}_1 = 1$, $\dot{z}_2 = \dot{z}_1/z_1 = 1/5$, $\dot{z}_3=2z_0\dot{z}_0=(2)(2)(1)=4$, $\dot{z}_4=\dot{z}_2 - \dot{z}_3=1/5 - 4$, and finally $\dot{y}=\dot{z}_4=-3.8$.
While this computation can be conducted in a single pass with evaluation of the original function $f$,
computing another derivative $\frac{\partial f}{\partial x_1}$ requires a separate pass.
Thus, forward mode is inefficient if the whole gradient of a function with many input variables is needed, e.g., the loss function of a high-dimensional model.
Reverse-mode AD applies the chain rule in the opposite direction. In the first pass, the original function and the associated intermediate variables $z_i$ are evaluated from input to output. In the second pass, the ``adjoint'' variables $\bar{z}_i\equiv\frac{\partial y}{\partial z_i}$ are initialized to zero and updated from output to input. In Fig. \ref{fig:cgraph}, 
$\bar{z}_4\pluseq\frac{\partial y}{\partial z_4}=1$,
$\bar{z}_3\pluseq\bar{z}_4\frac{\partial z_4}{\partial z_3}=-1$,
$\bar{z}_2\pluseq\bar{z}_4\frac{\partial z_4}{\partial z_2}=1$,
$\bar{z}_0\pluseq\bar{z}_3\frac{\partial z_3}{\partial z_0} = \bar{z}_3 (2z_0)=-4$,
$\bar{z}_1\pluseq\bar{z}_2\frac{\partial z_2}{\partial z_1} = \frac{\bar{z}_2}{z_2}=1/5$,
$\bar{z}_0\pluseq\bar{z}_1\frac{\partial z_1}{\partial z_0} = 1/5$,
and
$\bar{z}_{-1}\pluseq\bar{z}_1\frac{\partial z_1}{\partial z_{-1}}=1/5$.
Here, the `$\pluseq$' is the C-style increment operator, employed in order to observe the rule of total derivatives. (Note $\bar{z}_0$ is updated twice.)  
Finally, $\frac{\partial f}{\partial x_1}=\bar{x}_1=\bar{z}_{-1}=0.2$ and $\frac{\partial f}{\partial x_2}=\bar{x}_2 = \bar{z}_0 = -3.8$.
Hence reverse-mode AD generalizes the backpropagation algorithm and computes the whole gradient $\nabla f$ in a single backward pass, at the expense of keeping intermediate variables.

Deep learning software can be categorized by the way they build computational graphs. In Theano and TensorFlow, the user needs to construct a \emph{static} computational graph using a specialized mini-language before executing the model fitting process, and the graph cannot be modified throughout the execution. This static approach has performance advantage since there is room for optimizing the graph structure. Its disadvantage is the limited expressiveness of computational graphs and AD. On the other hand, PyTorch employs \emph{dynamic} computational graphs, for which the user describes the model as a regular program for (forward) evaluation of the loss function. Intermediate values and computation trace are recorded in the forward pass, and the gradient is computed by parsing the recorded computation backwards.
The advantage of this dynamic graph construction is the expressiveness of the model: in particular, recursion is allowed in the loss function definition. For example, recursive models 
such as $f(x) = f(x/2)$ if $x>1$ and $x$ otherwise
are difficult to describe using a static graph but easy with a dynamic one.
The downside is slower evaluation due to function call overheads.


\subsection{Case study: PyTorch versus TensorFlow}\label{sec:mcpi_torch}
In this section, we illustrate how simple it is to write a statistical computing code on multi-device HPC environments using modern deep learning libraries. 
We compare PyTorch and TensorFlow code written in Python, which computes a Monte Carlo estimate of the constant $\pi$.
The emphasis is on readability and flexibility, i.e., how small a modification is needed to run the code written for a single-CPU node on a multi-GPU node and a multi-node system.

Listing \ref{code:torch_mcpi} shows the code for Monte Carlo estimation of $\pi$ using PyTorch. Even for those who are not familiar with Python, the code should be quite readable. The main workhorse is  function \texttt{mc\_pi()} (Lines 14--21), which generates a sample of size $n$ from the uniform distribution on the unit square $[0, 1]^2$ and compute the proportion of the points that fall inside the quarter circle of unit radius centered at the origin. Listing \ref{code:torch_mcpi} is a fully executable program. It uses \texttt{torch.distributed} interface with an MPI backend (Line 3). An instance of the program of Listing \ref{code:torch_mcpi} is attached to a device and is executed as a ``process''. Each process is given its identifier (rank), which is retrieved in Line 6. The total number of processes is known to each process via Line 7.
After the proportion of the points in the quarter-circle is computed in Line 22,
each process gathers the sum of the means computed from all the processes in Line 25 (this is called the all-reduce operation; see Section \ref{sec:environ:prelim}). Line 27 divides the sum by the number of processes, yielding a Monte Carlo estimate of $\pi$ based on the sample size of $n \times \text{(number of processes)}$.

We have been deliberately ambiguous about the ``devices.'' Here, a CPU core or a GPU is referred to as a device. 
Listing \ref{code:torch_mcpi} assumes  the environment is a workstation with one or more GPUs, and the backend MPI is CUDA-aware. A CUDA-aware MPI, e.g., OpenMPI \citep{openmpi}, allows data to be sent directly from a GPU to another GPU through the MPI protocols.
Data transfer between modern GPUs does not go through CPU \citep{lee2017large}.
Lines 9 --10 specify that the devices to use in the program are GPUs. 
%
For example, suppose the workstation has four GPUs, say device 0 through 3.
A likely scenario for carrying out the all-reduce operation in Line 25 is to transfer the estimated $\pi$ in device 1 (computed in Line 22, which is parallelized) to device 0, where the two estimates are added. 
At the same time, the estimate in device 3 is passed to device 2 and then added with another estimate there. After this step, the sum in device 2 is sent to device 0 to compute the final sum. This sum is broadcast to all the other devices to replace the local estimates.
(The actual behavior may be slightly different from this scenario depending on the specific implementation of MPI.)
%
If the environment is a cluster with multiple CPU nodes (or even a single node), 
then communication between nodes or CPU cores through high-speed interconnect replaces the inter-GPU communication. 
At the code level, all we need to do is change Line 10 to \texttt{device = 'cpu'}.
The resulting code runs on a cluster seamlessly as long as the MPI for the cluster is properly installed.

\begin{lstfloat}[t]
\begin{lstlisting}[label=code:torch_mcpi,caption=Distributed Monte Carlo estimation of $\pi$ using PyTorch]
# import packages
import torch.distributed as dist
import torch
dist.init_process_group('mpi')  # initialize MPI

rank = dist.get_rank()          # device id
size = dist.get_world_size()    # total number of devices

# select device 
device = 'cuda:{}'.format(rank) # or simply 'cpu' for CPU computing
# select GPU based on rank.
if device.startswith('cuda'): torch.cuda.set_device(rank)

def mc_pi(n): 
    # this code is executed on each device.
    # generate n samples from Unif(0, 1) for x and y
    x = torch.rand((n), dtype=torch.float64, device=device)
    y = torch.rand((n), dtype=torch.float64, device=device)
    # compute local estimate of pi in float64. 
    # type conversion is necessary, because (x ** 2 + y ** 2 < 1)
    # results in unsigned 8-bit integer. 
    pi_hat = torch.mean((x**2 + y**2 <1).to(dtype=torch.float64))*4
    # sum of the estimates across processes
    #  is stored in-place in 'pi_hat', overwriting its original value.
    dist.all_reduce(pi_hat) 
    # the final estimate of pi, computed on each process
    return pi_hat / size

if __name__ == '__main__':
    n = 10000
    pi_hat = mc_pi(n)
    print("Pi estimate based on {} Monte Carlo samples across {} processes.".format(n * size, size))
    if rank == 0:
        print(pi_hat.item())
\end{lstlisting}
\end{lstfloat}

In TensorFlow, however, a separate treatment of multi-GPU and cluster settings is almost necessary. The code for multi-GPU setting is similar to Listing \ref{code:torch_mcpi} and is given in Appendix \ref{sec:mcpi_tf}. 
%
%
In a cluster setting, unfortunately, it is extremely difficult to reuse the multi-GPU code. 
If direct access to individual compute nodes is available, that information can be used to run the code distributedly, albeit not intuitively.  
However, in HPC environments where computing jobs are managed by job schedulers, we often do not have direct access to the compute nodes. 
The National Energy Research Scientific Computing Center (NERSC), the home of the 16th most powerful supercomputers in the world (as of June 2020), advises that gRPC, the default inter-node communication method of TensorFlow, is very slow on tightly-coupled nodes, thus recommends a direct use of MPI \citep{nersc:tf}.
Using MPI with TensorFlow requires an external library called Horovod and a substantial modification of the code, as shown in Listing \ref{code:tf_mcpi_horovod}.
This is a sharp contrast to Listing \ref{code:torch_mcpi}, where essentially the same PyTorch code can be used in both multi-GPU and multi-node settings.

Due to the reasons stated in Section \ref{sec:ad}, we employ PyTorch in the sequel to implement the highly parallelizable algorithms of Section \ref{sec:algo} in a multi-GPU node and a cluster on a cloud, as it allows simpler code that runs on various HPC environments with a minimal modification. (In fact, this modification can be made automatic through a command line argument.)

\begin{lstfloat}[ht!]
\begin{lstlisting}[label=code:tf_mcpi_horovod,caption=Monte Carlo estimation of $\pi$ for TensorFlow on multiple nodes using Horovod]
import tensorflow as tf
import horovod.tensorflow as hvd

# initialize horovod
hvd.init()
rank = hvd.rank()

# without this block, all the processes try to allocate 
# all the memory from each device, causing out of memory error.
devices = tf.config.experimental.list_physical_devices("GPU")
if len(devices) > 0:
    for d in devices:
        tf.config.experimental.set_memory_growth(d, True)

# select device
tf.device("device:gpu:{}".format(rank)) # tf.device("device:cpu:0") for CPU

# function runs in parallel with (graph computation/lazy-evaluation)
# or without (eager execution) the line below
@tf.function 
def mc_pi(n):
    # this code is executed on each device
    x = tf.random.uniform((n,), dtype=tf.float64)
    y = tf.random.uniform((n,), dtype=tf.float64)
    # compute local estimate for pi and save it as 'estim'.
    estim = tf.reduce_mean(tf.cast(x**2 + y ** 2 <1, tf.float64))*4
    # compute the mean of 'estim' over all the devices
    estim = hvd.allreduce(estim)
    return estim

if __name__ == '__main__':
    n = 10000
    estim = mc_pi(n)
    # print the result on rank zero
    if rank == 0:
        print(estim.numpy())
\end{lstlisting}
\end{lstfloat}

\section{Highly Parallelizable Algorithms}\label{sec:algo}
In this section, we discuss some easily parallelizable optimization algorithms useful for fitting high-dimensional statistical models, assuming that data are so large that they have to be stored distributedly. These algorithms can benefit from the distributed-memory environment by using relatively straightforward operations, via distributed matrix-vector multiplication and independent update of variables.

\subsection{MM algorithms}\label{sec:mm}
The MM principle \citep{lange2000optimization,lange2016mm}, 
where ``MM'' stands for either majorization-minimization 
or minorization-maximization,  is a useful tool for constructing parallelizable optimization algorithms. 
In minimizing an objective function $f(x)$ iteratively, for each iterate we consider a surrogate function $g(x|x^n)$ satisfying two conditions: the tangency condition $f(x^n) = g(x^n | x^n)$ and the domination condition $f(x) \le g(x | x^n)$ for all $x$. Updating $x^{n+1} = \arg\min_x g(x | x^n)$ guarantees that $\{f(x^n)\}$ is a nonincreasing sequence:
$$
f(x^{n+1}) \le g(x^{n+1} | x^n) \le g(x^n | x^n) = f(x^n).
$$
In fact, full minimization of $g(x | x^n)$ is not necessary for the descent property to hold; merely decreasing it is sufficient. 
For instance, it can be shown that the EM algorithm \citep{dempster1977maximum} is obtained by applying the MM principle to 
to the observed-data log likelihood
and Jensen's inequality.
(See \citet{wu2010mm} for more details about the relation between MM and EM.)
 
MM updates are usually designed to make a nondifferentiable objective function smooth, linearize the problem, or avoid matrix inversions by a proper choice of a surrogate function. MM is naturally well-suited for parallel computing environments, as we can choose a separable surrogate function and update variables independently.  
For example, 
when maximizing loglikelihoods, a term involving summation inside the logarithm $\log(\sum_{i=1}^p u_i)$, $u_i > 0$, often arises.
By using Jensen's inequality, this term can be minorized and separated as
$$
\log\left(\sum_{i=1}^p u_i\right) \ge \sum_{i=1}^p \frac{u_i^n}{\sum_{j=1}^p u_j^n} \log \left( \frac{\sum_{j=1}^p u_j^n}{u_i^n} u_i\right) =  \sum_{i=1}^p \left(\frac{u_i^n}{\sum_{j=1}^p u_j^n}\right) \log u_i + c_n,
$$
where $u_i^n$'s are constants and $c_n$ is a constant only depending on $u_i^n$'s.
Parallelization of MM algorithms on a single GPU using separable surrogate functions is extensively discussed in \citet{zhou2010graphics}. Separable surrogate functions are especially important in distributed HPC environments, e.g. multi-GPU systems.  

\subsection{Proximal gradient method}\label{sec:proxgrad}
The proximal gradient method 
extends the gradient descent method, and
deals with minimization of sum of two extended real-valued convex functions, i.e., 
\begin{equation}\label{eqn:sumtwo}
\min_x f(x) + g(x).
\end{equation}
Function $f$ is possibly nondifferentiable, while $g$ is continuously differentiable. 

We first define the proximity operator of $f$: 
$$
\mathbf{prox}_{\lambda f} (y) = \arg\min_x \left\{ f(x) + {\textstyle\frac{1}{2 \lambda}} \|x - y \|_2^2 \right\},\;\; \lambda > 0
$$
For many functions their proximity operators take closed forms. 
We call such functions ``proximable''. 
For example, consider the $0/\infty$ indicator function $\delta_C (x)$
of a closed convex set $C$,
i.e., $\delta_C (x) = 0$ if $x \in C$, and $+\infty$ otherwise.
The corresponding proximity operator is the Euclidean projection onto $C$: $P_C (y) = \arg\min_{x \in C} \|y - x\|_2$. 
For many sets, e.g., nonnegative orthant, $P_C$ is simple to compute. 
Also note that
the proximity operator of the $\ell_1$-norm $\lambda \| \cdot \|_1$ is the soft-thresholding operator:
$[\mathcal{S}_\lambda (y)]_i := \mathrm{sign}(y_i) (|y_i| - \lambda)_+$.

Now we proceed with the proximal gradient method for minimization of $h(x) = f(x) + g(x)$. Assume $g$ is convex and has an $L$-Lipschitz gradient, i.e., $\|\nabla g(x) - \nabla g(y)\|_2 \le L\|x - y\|_2$ for all $x$, $y$ in the interior of its domain, and $f$ is lower-semicontinuous, convex, and  proximable. The $L$-Lipschitz gradients naturally result in the following surrogate function that majorizes $h$:
\begin{align*}
h(x) & \le f(x) + g(x^n) + \langle \nabla g(x^n), x - x^n \rangle +        {\textstyle\frac{L}{2}} \|x - x^n\|_2^2 \\
     & = f(x) + g(x^n) + {\textstyle\frac{L}{2}} \left\|x - x^n + {\textstyle\frac{1}{L}} \nabla g(x^n)\right\|_2^2 - {\textstyle\frac{1}{2L}} 
     \left\|\nabla g(x^n)  \right\|_2^2 =: p(x|x^n) .
\end{align*}
Minimizing $p(x|x^n)$ with respect to $x$ results in the iteration:
\begin{align}\label{eqn:proxgrad}
x^{n+1} = \mathbf{prox}_{\gamma_n f} \left(x^n - \gamma_n  \nabla g(x^n)\right), \;\; \gamma_n \in \left(0,  {1}/{L}\right].
\end{align}
If $f\equiv 0$, then iteration \eqref{eqn:proxgrad} reduces to the conventional gradient descent.
This iteration guarantees a nonincreasing sequence of $h(x^n)$ by the MM principle.
Proximal gradient method also has an interpretation of forward-backward operator splitting, and the step size $\gamma_n \in \left( 0, {2}/{L}\right)$ guarantees convergence 
\citep{combettes2011proximal,combettes2018monotone}.
If $f(x) = \delta_C (x)$, then the corresponding algorithm is called the projected gradient method. If $f(x) = \lambda \|x\|_1$, then 
it
is the iterative shrinkage-thresholding algorithm \citep[ISTA,][]{Beck:SiamJournalOnImagingSciences:2009}. 

For many functions $f$, the update \eqref{eqn:proxgrad} is simple and easily parallelized,
thus the algorithm is suitable for HPC. 
For example, in the soft-thresholding operator above all the elements are independent.
If $f(x)= -a \log x$, then 
\begin{align}\label{eqn:pet}
\mathbf{prox}_{\gamma f} (y) = 
\big(y + \sqrt{y^2 + 4 \gamma a}\big)/2,
\end{align}
which is useful for the PET example in Section \ref{sec:num}.
The gradient $\nabla g$ in update \eqref{eqn:proxgrad} can also be computed in parallel. In many models the fitting problem takes the form of \eqref{eqn:sumtwo} with $g(x)=\frac{1}{m}\sum_{i=1}^m\ell(a_i^Tx)$, where $\ell$ is a loss function and $a_i \in \mathbb{R}^p$ is the $i$th observation.
Collect the latter into a data matrix $A \in \mathbb{R}^{m \times p}$. If $m \gg p$, then split it by the row as $A = [A_{[1]}^T, A_{[2]}^T, \cdots, A_{[d]}^T]^T$, where blocks $A_{[k]}$ 
are distributed over $d$ devices. If the current iterate of the parameter $x^n$ is known to each device, then the local gradient $\nabla g_i(x^n)=\ell'(a_i^Tx)a_i$ can be computed from $A_{[k]}$ independently. The full gradient $\nabla g(x^n)$ can be computed then by averaging $\nabla g_i(x^n)$.
In the MPI terminology of Section \ref{sec:environ:prelim}, 
a distributed-memory proximal gradient update consists of the following steps:
1) broadcast $x^{n}$;
2) compute the local gradient $\nabla g_i(x^n)$ in each device;
3) reduce the local gradients to compute the full gradient $\nabla g(x^n)$ in the master device;
4) update $x^{n+1}$.
%
If $g$ is not separable in observations, splitting the data matrix by column 
may be useful 
(Section \ref{sec:cox}).

See \citet{parikh2014proximal} for a thorough review and distributed-memory implementations, and \citet{polson2015proximal} for a statistically oriented review.

\subsection{Primal-dual methods}\label{sec:pdhg}
Primal-dual methods introduce an additional dual variable $y$ (where $x$ is the primal variable) in order to deal with a larger class of problems. 
Consider the problems of the form $h(x) = f(K x) + g(x)$, where $K \in \mathbb{R}^{l \times p}$. We further assume that 
$f$ and $g$ 
are lower semicontinuous, convex, and proper 
(i.e., not always $\infty$)
functions.
Even if $f$ is proximable, the proximity operator for $f(K \cdot)$ is not easy to compute. 
The conjugate of $f$ is defined as
$f^*(y) = \sup_x \langle x, y \rangle - f(x)$. 
It is known that $f^{**} = f$, 
so $f(Kx) = f^{**}(Kx) = \sup_y \langle Kx, y \rangle - f^* (y)$. Then  the minimization problem $\inf_x f(Kx) + g(x)$ is equivalent to the saddle-point problem 
$$
\inf_x \sup_y \langle Kx, y \rangle + g(x)  - f^*(y)
,
$$
for which a solution $(\hat{x}, \hat{y})$ exists under mild conditions.

A widely known method for solving this saddle-point  problem in the statistical literature is the ADMM \citep{xue2012positive,ramdas2016fast,zhu2015augmented,lee2017large,gu2018admm}, 
whose
update is given by:
\begin{subequations}\label{eqn:admm}
\begin{align}
x^{n+1} &= \arg\min_x g(x) + (t/2) \|Kx - \tilde{x}^n + (1/t) y^n\|_2^2 \label{eqn:admm:inner}\\
\tilde{x}^{n+1} &= \mathbf{prox}_{(1/t)f} (Kx^{n+1} + (1/t) y^n) \\
y^{n+1} &= y^n + t(Kx^{n+1} - \tilde{x}^{n+1}).
\end{align}
\end{subequations}
If $g$ is separable, i.e., $g(x) = \sum_{k=1}^d g_k(x)$, then
consensus optimization \citep[Chap. 7]{Boyd:FoundationsAndTrendsInMachineLearning:2010} applies ADMM to distributed copies of variables $x_k=x$ to  minimize $h(x) = f(z) + \sum_{k=1}^d g_k(x_k)$ subject to $x_k=x$ and $Kx_k=z$ for each $k$:
\begin{subequations}\label{eqn:consensus}
\begin{align}
x^{n+1}_k &= \arg\min_{x_k} g_k(x_k) + {\textstyle\frac{t}{2}} \|K x_k - \tilde{x}^n + {\textstyle\frac{1}{t}}y_k^n\|_2^2 + {\textstyle\frac{t}{2}}\|x_k - x^n + {\textstyle\frac{1}{t}}w_k^n\|_2^2 \label{eqn:consensus:inner}\\
\tilde{x}^{n+1} &= \mathbf{prox}_{(dt)^{-1}f} \big(\textstyle\frac{1}{d}\sum_{k=1}^d (K x_k^{n+1} + \frac{1}{t} y_k^n ) \big) \label{eqn:consensus:2}\\
y^{n+1}_k &= y^n_k + t(Kx^{n+1}_k - \tilde{x}^{n+1}),
\quad 
w^{n+1}_k = w^n_k + t(x^{n+1}_k - x^{n+1})
.
\label{eqn:consensus:3}
\end{align}
\end{subequations}
A distributed-memory implementation will iterate the following steps:
1) for each device $k$, solve \eqref{eqn:consensus:inner} in parallel;
2) gather local solutions $x_k^{n}$ in the master device;
3) compute \eqref{eqn:consensus:2};
4) broadcast $\tilde{x}^{n+1}$;
5) compute \eqref{eqn:consensus:3}.

Nonetheless, 
neither update \eqref{eqn:admm:inner} 
nor \eqref{eqn:consensus:inner}
results in
a proximity operator, since the quadratic term is not spherical. 
This inner optimization problem is often nontrivial to solve. 
In the simplest case of linear regression, 
$g$ is quadratic and
\eqref{eqn:admm:inner} involves solving a (large) linear system 
whose time complexity is cubic in the 
dimension $p$ of the primal variable $x$.

PDHG
avoids inner optimization 
via the following iteration:
\begin{subequations}\label{eqn:cp}
\begin{align}
y^{n+1} &= \mathbf{prox}_{\sigma f^*}(y^n + \sigma K \bar{x}^n) \label{eqn:cp:dualprox} \\
x^{n+1} &= \mathbf{prox}_{\tau g} (x^n - \tau K^T y^{n+1}) \label{eqn:cp:primalprox} \\
\bar{x}^{n+1} &= 2x^{n+1}- x^n, \label{eqn:cp:extrapolate} 
\end{align}
\end{subequations}
where 
$\sigma$ and $\tau$ are step sizes.
If $f$ is proximable, so is $f^*$, since $\mathbf{prox}_{\gamma f^*}(x) = x - \gamma \mathbf{prox}_{\gamma^{-1} f}(\gamma^{-1} x)$ by Moreau's decomposition.
This method has been analyzed using monotone operator theory \citep{Condat:JournalOfOptimizationTheoryAndApplications:2012,Vu2013,ko2019easily}.
Convergence of iteration \eqref{eqn:cp} is guaranteed if $\sigma \tau \|K\|_2^2 < 1$, where $\|M\|_2$ is the spectral norm of matrix $M$. 
If $g$ has an $L$-Lipschitz gradient, then the proximal step \eqref{eqn:cp:primalprox} can be replaced by a gradient step 
$$
x^{n+1} = x^n - \tau(\nabla g(x^n) + K^T y^{n+1}).
$$
PDHG algorithms are also highly parallelizable as long as the involved proximity operators are easy to compute and separable. 
No 
inner optimization
is involved in iteration \eqref{eqn:cp} and only matrix-vector multiplications appear.
The distributed computation of gradient in Section \ref{sec:proxgrad} can be used for the gradient step.
A hybrid of PDHG and ADMM has recently been proposed \citep{ryu2018splitting}.

\subsection{Parallel coordinate descent and stochastic approximation}
Coordinate descent methods apply vector-to-scalar maps $T_i:\mathbb{R}^p\to\mathbb{R}: x=(x_1,\dotsc, x_i, \dotsc, x_p) \mapsto \arg\min_{x_i'} h(x_1,\dotsc,x_i',\dotsc,x_p)$ defined for each coordinate $i$ successively to minimize $h(x)$.
The most well-known variant is the cyclic or Gauss-Seidel version. 
If we denote the $j$th elementary unit vector in $\mathbb{R}^p$ by $e_j$, then the update rule is $x^{n+1} = \sum_{j\neq i}x^n_je_j + T_i(x)e_i$ where $i = (n - 1 \mod p) + 1$, which possesses the descent property.
The parallel or Jacobi update reads $x^{n+1} = \sum_{j=1}^p T_j(x)e_j$. Obviously, if $h$ is separable in variables, i.e., $h(x)=\sum_{j=1}^p h_j(x_j)$, this minimization strategy will succeed. 
Other variants are also possible, such as randomizing the cyclic order, or updating a subset of coordinates in parallel at a time. The ``argmin'' map $T_i$ can also be relaxed, e.g., by a prox-linear map $x\mapsto \arg\min_{x_i'} \langle \frac{\partial g}{\partial x_i}|_{x_i}, x_i' - x_i \rangle + \frac{1}{2\gamma_i}\|x_i'-x_i\|_2^2 + f(x)$ if $h$ has a structure of $h=f + g$ and only $g$ is differentiable \citep{tseng2009}. See \citet{wright2015} for a recent review.

If $p$ is much larger than 
$\tau$, the number of devices, 
then choosing a subset of coordinates with size comparable to $\tau$ would reduce the complexity of an iteration. \citet{richtarik2016optimal,richtarik2016parallel} consider sampling a random subset and study the effect of the sampling distribution on the performance of parallel prox-linear updates, deriving optimal distributions for certain cases. In particular, the gain of parallelization is roughly proportional to the degree of separability $p/\omega$, where $\omega = \max_{J \in \mathcal{J}}|J|$ if $h(x) = \sum_{J\in\mathcal{J}}h_J(x)$ for a finite collection of nonempty subsets of $\{1,\dotsc, p\}$ and $h_J$ depends only on coordinates $i \in J$. For example, if $A=[a_1^T, \dotsc, a_m^T]^T\in\mathbb{R}^{m\times p}$ is the data matrix for ordinary least squares, then 
$\omega$ equals to the maximum number of nonzero elements in the rows, or equivalently $\omega = \max_{i=1,\dots,n}\|a_i\|_0$.

For gradient-descent type methods, stochastic approximation \citep[][see \citet{LaiYuan20SAReview} for a recent review]{RobbinsMonro51StochApprox} has gained wide popularity under the name of \emph{stochastic gradient descent} or SGD. 
The main idea is to replace the gradient of the expected loss by its unbiased estimator. For instance, as in the penultimate paragraph of Section \ref{sec:proxgrad}, if $g(x)=\frac{1}{m}\sum_{i=1}^m\ell(a_i^T x)$, and $f\equiv 0$, then $\ell'(a_i^T x)a_i$ is an unbiased estimator of $\nabla g(x)$ under the uniform distribution on the sample indices $\{1, \dotsc, m\}$. The update rule is then $x^{n+1} = x^n - \gamma_n \ell'(a_i^T x^n)a_i$ for some randomly chosen $i$.
SGD and its variants \citep{defazio2014,johnson2013} are main training methods in most deep learning software, since 
the sample size $m$ needs to be extremely large 
to properly train deep neural networks.
The idea of using an unbiased estimator of the gradient has been extended to the proximal gradient \citep{Nitanda2014,xiao2014proximal,atchade2017,rosasco2019} and PDHG \citep{chen2014optimal,ko2019easily,ko2019optimal} methods.
In practice, it is standard to use a minibatch or a random subset of the sample for each iteration, and the arbitrary sampling paradigm of \citet{richtarik2016optimal,richtarik2016parallel} for parallel coordinate descent has been extended to minibatch SGD \citep{gower2019,qian2019} and PDHG \citep{chambolle2018}.

\section{Distributed matrix data structure for PyTorch}\label{sec:distmat}
For the forthcoming examples and potential future uses in statistical computing, we propose the package \texttt{dist\_stat} built on PyTorch. It consists of two submodules, \texttt{distmat} and \texttt{application}. The submodule \texttt{distmat} implements a simple distributed matrix data structure, and the submodule \texttt{application} includes the code for the examples in Section \ref{sec:num} using \texttt{distmat}. In the data structure \texttt{distmat}, each process, enumerated by its rank, holds a contiguous block of the full data matrix by rows or columns, which may be sparse. If multiple GPUs are involved, each process controls the GPU whose index matches the process rank. The blocks are assumed to have equal sizes. For notational simplicity, we indicate the dimension to split by a pair of square brackets: if a $[100] \times 100$ matrix is split over four processes, the rank-0 process keeps the first 25 rows of the matrix,  the rank-1 process takes the next 25 rows, and so on. For the sake of simplicity, we always assume that the dimension to split is a multiple of the number of processes. 
The code for \texttt{dist\_stat} is available at \url{https://github.com/kose-y/dist\_stat}.
A proper backend setup for a cloud environment is explained in Appendix \ref{sec:cloud}.

In \texttt{distmat}, unary elementwise operations such as exponentiation, square root, absolute value, and logarithm of matrix entries are implemented in an obvious manner. 
Binary elementwise operations such as addition, subtraction, multiplication, division are implemented in a similar manner to R's vector recycling:
if two matrices of different dimensions are to be added together, say one is 
$3 \times 4$
and the other is 
$4 \times 1$,
the latter matrix is expanded to a 
$3 \times 4$
matrix with the column repeated four times. 
Another example is adding a 
$1 \times 4$
matrix and a 
$4 \times 1$
matrix. 
The former is expanded to a 
$4 \times 3$
matrix by repeating the row four times, and the latter to a 
$4 \times 3$
matrix by repeating the column three times. 
Application of this recycling rule is in accordance with the broadcast semantics of PyTorch.

%
Distributed matrix multiplication requires some care.
Suppose we multiply a $p \times r$ matrix $A$ and an $r \times q$ matrix $B$.
If matrix $B$ is tall and split by \emph{row} into $[B_{[1]}, \dotsc, B_{[T]}]^T$ and distributed among $T$ processes, where $B_{[t]}$ is the $t$-th row block of  $B$. 
If matrix $A$ is split in the same manner, 
a natural way to compute the product $AB$ is 
for each process $t$ to \emph{gather} (see Section \ref{sec:environ:prelim}) all $B_{[1]}, \dotsc, B_{[T]}$ to create a copy of $B$ and compute the row block $A_{[t]}B$ of $AB$.
On the other hand, if matrix $A$ is wide and split by \emph{column} into $[A^{[1]}, \dotsc, A^{[T]}]$, where $A^{[t]}$ is the $t$-th column block of  $A$, then
each process will compute the local multiplication $A^{[t]} B_{[t]}$.
The product $AB = \sum_{t=1}^T A^{[t]} B_{[t]}$ is computed by a \emph{reduce} or \emph{all-reduce} operation of Section \ref{sec:environ:prelim}.
These operations are parallelized as outlined in Section \ref{sec:mcpi_torch}.
The distribution scenarios considered in \texttt{distmat} are collected in Table \ref{tab:mm}.
Each matrix
can be either \emph{broadcast} ($p \times r$ for $A$), row-distributed ($[p] \times r$), or column-distributed ($p \times [r]$). 
Since broadcasting both matrices does not require any distributed treatment in multiplication, there remain eight possible combinations of the input.
For each combination,
the output may involve more than one configurations.
If an outer dimension (either $p$ or $q$ but not both) is distributed, the $p \times q$ output $AB$ is distributed along that dimension (scenarios 4, 8, 11). 
If both dimensions are split, then there are two possibilities of $[p] \times q$ and $p \times [q]$ (scenarios 2, 3).
%
Splitting of the inner dimension $r$ does not affect the distribution of the output unless it is distributed in both $A$ and $B$ (scenarios 1, 9, 10).
%
Otherwise, we
consider all the possible combinations in the output: broadcast, split by rows, and split by columns (scenarios 5, 6, 7).

The \texttt{distmat.mm()} function implements the 11 scenarios of Table \ref{tab:mm}
using the PyTorch function \texttt{torch.mm()} 
for within-process matrix multiplication
and the collective communication directives (Section \ref{sec:environ:prelim}). 
Scenarios 3, 6, 8, 10, and 11 are implemented using the transpositions of input and output matrices for scenarios 2, 7, 1, 9, and 4, respectively. 
Transposition costs only a short constant time, as it only `tags' to the original matrix that it is transposed. The data layout remains intact.
A scenario 
is automatically selected depending on the distribution of the input matrices. 
The class \texttt{distmat} has an attribute for determining if the matrix is distributed by row or column.
For scenarios 2, 3; 5, 6, and 7, which
share the same input structure,
additional keyword parameters are supplied to distinguish them and determine the shape of the output matrix.
%
The type of collective communication operation and the involved matrix block sizes roughly determine the communication cost of the computation. For example, an all-reduce is more expensive than a reduce.
The actual cost depends on the network latency, number of MPI messages sent, and sizes of the messages sent between processes, 
which are all system-dependent.


\begin{table}[]
\caption{Eleven distributed matrix multiplication scenarios of \texttt{distmat}. 
}\label{tab:mm}
\centering
\resizebox{\textwidth}{!}{%
\begin{tabular}{llllp{0.4\linewidth}l}
\hline
& $A$            & $B$             & $AB$           & Description    & \begin{tabular}[t]{@{}l@{}}Communication involved \\ (size of output) \end{tabular}                                       \\ \hline
1 & $[p] \times r$ & $[r] \times q$ & $[p] \times q$ & A wide matrix times a tall matrix.                                                                  
& 1 all-gather ($r \times q$) \\
2 & $[p] \times r$ & $r \times [q]$ & $[p] \times q$   & Outer product, may require a large amount of memory.
                                                                    &  1 all-gather ($r \times q$)                                                                                       \\
3 & $[p] \times r$ & $r \times [q]$ & $p \times [q]$   & Outer product, may require a large amount of memory. 
                                                                    &  1 all-gather ($r \times p$)                                                                                       \\
4 & $[p] \times r$ & $r \times q$ & $[p] \times q$ & A distributed matrix times a small, broadcast matrix.  & None \\ 
5 & $p \times [r]$ & $[r] \times q$ & $p \times q$ & Inner product, result broadcast.  Suited for inner product between two tall matrices.                                                                        &  1 all-reduce ($p \times q$)                                                                                        \\          
6 & $p \times [r]$ & $[r] \times q$ & $[p] \times q$ & Inner product, result distributed.                                                                                                                            & $T$ reductions ($p \times q/T$ each)                                                                                                  \\         
7 & $p \times [r]$ & $[r] \times q$ & $p \times [q]$ & Inner product, result distributed.                                                                                                                            & $T$ reductions ($q \times p/T$ each)                                                                                                  \\
8 & $p \times [r]$ & $r \times [q]$ & $p \times [q]$ & Multiply two column-distributed wide matrices                                                
& 1 all-gather ($p \times r$) \\
9 & $p \times [r]$ & $r \times q$ & $p \times q$   & A distributed matrix times a tall broadcast matrix. Intended for matrix-vector multiplications. &   1 all-reduce ($p \times q$)                                                                                      \\
10 & $p \times r$ & $[r] \times q$   & $p \times q$ & A tall broadcast matrix times a distributed matrix.  Intended for matrix-vector multiplications. &   1 all-reduce ($p \times q$) \\
11 & $p \times r$ & $r \times [q]$   & $p \times [q]$   & A small, broadcast matrix times a distributed matrix                                                                                               & None                                                                                                 \\ \hline
\end{tabular}
}
\end{table}

Listing \ref{code:distmat} demonstrates an example usage of \texttt{distmat}.
We assume that this program is run with four processes (\texttt{size} in Line 6 is \texttt{4}). Line 8 determines the device to use. If multiple GPUs are involved, the code selects one based on the rank of the process. Line 9 selects the GPU to use with PyTorch. This code runs on a system in which PyTorch is installed with 
a CUDA-aware MPI implementation.
The number of processes to be used can be supplied by a command-line argument (see Appendix \ref{sec:cloud}). 
Line 11 selects the data type and the device used for matrices. The \texttt{TType} (for ``tensor type'') of \texttt{torch.cuda.FloatTensor} indicates that single-precision GPU arrays are used, while  \texttt{DoubleTensor} employs double-precision CPU arrays. 
Then Line 12 creates a distributed $[4] \times 4$ matrix and initializes it to uniform $(0, 1)$ random numbers. 
This matrix is created once and initialized locally, and then distributed to all processes.
(For large matrices, \texttt{distmat} supports another creation mode that assembles matrix blocks from distributed processes.) 
%
%
Line 14 multiplies the two such matrices $A$ and $B$ to form a distributed matrix of size $[4] \times 2$.
Scenario 1 in Table \ref{tab:mm} is chosen by \texttt{distmat} to create the output $AB$.
Line 18 computes an elementwise logarithm of $1+AB$, 
in an elementwise fashion according to the recycling rule.
The local block of data residing in each process can be accessed by appending \texttt{.chunk} to the name of the distributed matrix, as in Lines 17 and 21.\footnote{%
Lines 17 and 21 do not guarantee printing in order (of ranks). 
They are printed on a first come, first served basis.%
}

Although the present implementation only deals with matrices, \texttt{distmat} can be easily extended to tensor multiplication, as long as the distributed multiplication scenarios are carefully examined as in Table \ref{tab:mm}. 
Creating communication-efficient parallel strategies that minimize the amount of communication between computing units
is an active area of research \citep{van1997summa,ballard2011minimizing,koanantakool2016communication}.
Communication-avoiding sparse matrix multiplication
has been utilized for sparse inverse covariance estimation \citep{koanantakool2018communication}.

\begin{lstfloat}[t!]
\begin{lstlisting}[label=code:distmat,caption=An example usage of the module \texttt{distmat}.]
import torch
from dist_stat import distmat
import torch.distributed as dist
dist.init_process_group('mpi')
rank = dist.get_rank()
size = dist.get_world_size()

device = 'cuda:{}'.format(rank) # or 'cpu' for CPU computing
if device.startswith('cuda'): torch.cuda.set_device(rank)

TType = torch.cuda.FloatTensor if device.startswith('cuda') else torch.DoubleTensor    # single precision for GPUs
A = distmat.distgen_uniform(4, 4, TType=TType) # create [4] x 4 matrix
B = distmat.distgen_uniform(4, 2, TType=TType) # create [4] x 2 matrix
AB = distmat.mm(A, B) # A * B, Scenario 1.
if rank == 0: # to print this only once
    print("AB = ")
print(rank, AB.chunk) # print the rank's portion of AB.
C = (1 + AB).log() # elementwise logarithm
if rank == 0:
    print("log(1 + AB) = ")
print(rank, C.chunk) # print the rank's portion of C.
\end{lstlisting}
\end{lstfloat}

\section{Examples}\label{sec:num}

In this section, we compare the performance of the optimization algorithms of Section \ref{sec:algo} on various HPC environments for the following four statistical computing examples using \texttt{distmat}: nonnegative matrix factorization (NMF), positron emission tomography (PET), multi-dimensional scaling (MDS),  
all of which were considered in \citet{zhou2010graphics},
and  $\ell_1$-regularized Cox proportional hazards regression for survival analysis. 
For the former three examples the focus is on scaling up the size of feasible problems from those about a decade ago.
%
For the last example, we focus on analyzing a real-world geonomic dataset of  size approximately equal to $200,000 \times 500,000$.
\subsection{Setup}\label{sec:num:setup}
%

We employed a local multi-GPU workstation and a virtual cluster consisted of multiple AWS EC2 instances for computing.
Table \ref{tab:setting} shows the setting of our HPC systems used for the experiments. 
For virtual cluster experiments, we utilized 1 to 20 of AWS \texttt{c5.18xlarge} instances with 36 physical cores with AVX-512 (512-bit advanced vector extension to the x86 instruction set) 
enabled in each instance through the CfnCluster resource manager. 
Network bandwidth of each \texttt{c5.18xlarge} instance was 25GB/s. A separate \texttt{c5.18xlarge} instance served as the ``master'' instance, which did not participate in computation by itself but managed the computing jobs over the 1 to 20 ``worker'' instances. Data and software for the experiments were stored in an Amazon Elastic Block Store (EBS) volume attached to this instance and shared among the worker instances via the network file system. Further details are given in Appendix \ref{sec:cloud}.
For GPU experiments, we used a local machine with two CPUs (10 cores per CPU) and eight Nvidia GTX 1080 GPUs. 
These are desktop GPUs, not optimized for double-precision.
All the experiments were conducted using PyTorch version 0.4 built on the Intel Math Kernel Library (MKL); 
the released code works for the versions up to 1.6.

We evaluated the objective function once per 100 iterations. For the comparison of execution time, the iteration was run for a fixed number of iterations, regardless of convergence. 
For comparison of different algorithms for the same problem, we iterated until $\frac{|f(\theta^n) - f(\theta^{n-100})|}{|f(\theta^n)| + 1} < 10^{-5}$.

For all the experiments, single-precision computation results on GPU agreed with  double-precision ones up to six significant digits, 
except for $\ell_1$-regularized Cox regression, where the PyTorch implementation of the necessary cumulative sum operation  caused numerical instability in some cases. 
Therefore all the experiments for Cox regression were carried out in double-precision. Extra efforts for writing a multi-device code were modest with \texttt{distmat}. Given around 1000 lines of code to implement basic operations for multi-device configuration in \texttt{distmat}, additional code for our four examples was less than 30 lines for each. 


\begin{table}[]
\caption{HPC environments for experiments}\label{tab:setting}
\centering
\begin{tabular}{lccc}
\hline
& \multicolumn{2}{c}{local node} & AWS c5.18xlarge \\
             & CPU                   & GPU             & CPU\\ \hline
Model        & Intel Xeon E5-2680 v2 & Nvidia GTX 1080 & \begin{tabular}{@{}c@{}}Intel Xeon \\ Platinum 8124M\end{tabular} \\
\# of cores     & 10                    & 2560            & 18\\
Clock        & 2.8 GHz                 & 1.6 GHz           & 3.0GHz \\
\# of entities        & 2                     & 8               & \begin{tabular}{@{}c@{}}2 (per instance) \\ $\times$ 1-20 (instances)\end{tabular} \\
Total memory & 256 GB                 & 64 GB            & 144 GB $\times$ 1--20 \\
Total cores  & 20                    & 20,480 (CUDA)          & 36 $\times$ 1--20 \\ \hline
\end{tabular}
\end{table}

\subsection{Scaling up examples in \citet{zhou2010graphics}}\label{sec:zhou}

\paragraph{Nonnegative matrix factorization}
NMF is a procedure that approximates a nonnegative data matrix $X \in \mathbb{R}^{m \times p}$ by a product of two low-rank nonnegative matrices, $V \in \mathbb{R}^{m \times r}$ and $W \in \mathbb{R}^{r \times p}$. In a simple setting, NMF minimizes
$f(V, W) =  \|X - VW\|_\mathrm{F}^2$,
where $\| \cdot \|_\mathrm{F}$ denotes the Frobenius norm. 
Applying the MM principle to recover the famous multiplicative algorithm due to \citet{lee1999learning,lee2001algorithms} is discussed in \citet[Sect. 3.1]{zhou2010graphics}. 
Alternatively, the alternating projected gradient (APG) method \citep{lin2007projected} introduces ridge penalties to minimize
$f(V, W; \epsilon) =  \|X - VW\|_\mathrm{F}^2 + \frac{\epsilon}{2} \|V\|_\mathrm{F}^2 + \frac{\epsilon}{2} \|W\|_\mathrm{F}^2$.
Then the APG iteration is given by
\begin{align*}
V^{n+1} &= P_+ \left((1 - \sigma_n \epsilon) V^n - \sigma_n (V^n W^n (W^n)^T - X (W^n)^T) \right) \\
W^{n+1} &= P_+ \left((1 - \tau_n \epsilon) W^n - \tau_n ((V^{n+1})^T V^{n+1} W^n - (V^{n+1})^TX ) \right),
\end{align*}
where $P_+$ denotes the projection onto the nonnegative orthant; $\sigma_n$ and $\tau_n$ are step sizes. 
Convergence is guaranteed if $\epsilon >0$, $\sigma_n \le 1/(2 \|W^n (W^n)^T + \epsilon I\|_\mathrm{F}^2)$, and $\tau_n \le 1/(2 \|(V^n)^T V^n + \epsilon I\|_\mathrm{F}^2)$. 
APG has an additional advantage of avoiding creation of subnormal numbers over the multiplicative algorithm (see Appendix \ref{sec:numextra}).
Table \ref{tab:nmf3} compares the performance of APG between single-machine multi-GPU and multi-instance virtual cluster settings. Synthetic datasets of sizes [10,000] $\times$ 10,000 and [200,000] $\times$ 200,000 were created and distributed. For reference, the dimension used in \citet{zhou2010graphics} is $2429 \times 361$. Multi-GPU setting achieved up to 4.14x-speedup over a single CPU instance if the dataset was small, but could not run the larger dataset. The cluster in a cloud was scalable with data, running faster with more instances, yielding up to 4.10x-speedup over the two-instance cluster.

\paragraph{Positron emission tomography}

\begin{figure}[t!]
\centering
\begin{subfigure}[t]{0.22\textwidth}
\includegraphics[width=\textwidth]{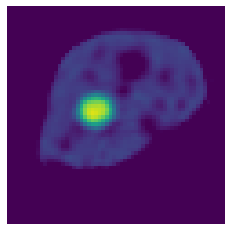}
\caption{PDHG, $\rho=0$}
\end{subfigure}
\begin{subfigure}[t]{0.22\textwidth}
\includegraphics[width=\textwidth]{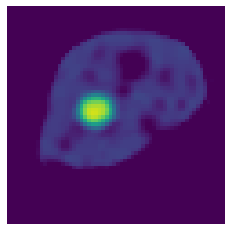}
\caption{$\rho=10^{-3}$}
\end{subfigure}
\begin{subfigure}[t]{0.22\textwidth}
\includegraphics[width=\textwidth]{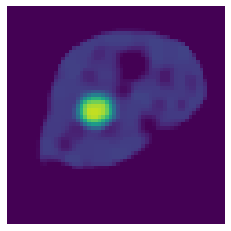}
\caption{$\rho=10^{-2.5}$}
\end{subfigure}
\begin{subfigure}[t]{0.22\textwidth}
\includegraphics[width=\textwidth]{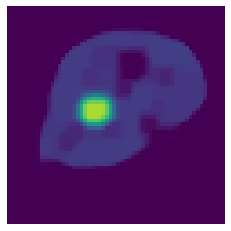}
\caption{$\rho=10^{-2}$} 
\end{subfigure}\\
\begin{subfigure}[t]{0.22\textwidth}
\includegraphics[width=\textwidth]{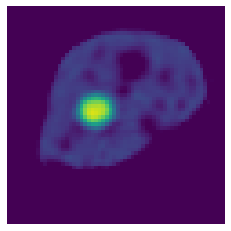}
\caption{SPDHG, $\rho=0$}
\end{subfigure}
\begin{subfigure}[t]{0.22\textwidth}
\includegraphics[width=\textwidth]{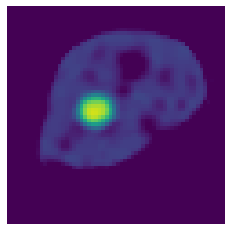}
\caption{$\rho=10^{-3}$}
\end{subfigure}
\begin{subfigure}[t]{0.22\textwidth}
\includegraphics[width=\textwidth]{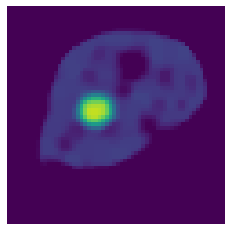}
\caption{$\rho=10^{-2.5}$}
\end{subfigure}
\begin{subfigure}[t]{0.22\textwidth}
\includegraphics[width=\textwidth]{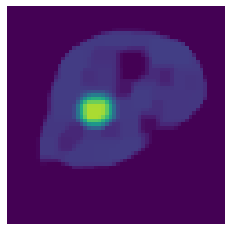}
\caption{$\rho=10^{-2}$} 
\end{subfigure}\\
\caption{Reconstruction of the XCAT phantom with a TV penalty with regularization parameter $\rho$, using deterministic (top row) and stochastic (bottom) PDHG.}
\label{fig:pet_l1_xcat}
\end{figure}

PET reconstruction is essentially a deconvolution problem of estimating the intensities of radioactive biomarkers from their line integrals,
which can be
posed as maximizing the Poisson loglikelihood
$L(\lambda) = \sum_{i=1}^d [ y_i \log (\sum_{j=1}^pe_{ij}\lambda_j ) - \sum_{j=1}^p e_{ij} \lambda_j]$. Here $y_i$ is the observed count of photons arrived coincidentally at 
detector pair $i$.
Emission intensities $\lambda = (\lambda_1, \cdots, \lambda_p)$ are to be estimated, and $e_{ij}$ is the probability that detector pair $i$ detects an emission form pixel location $j$, which dependes on the geometry of the detector configuration. 
We consider a circular geometry for two-dimensional imaging.
Adding a ridge-type penalty of $- (\mu/2) \|D \lambda\|_2^2$
to enhance spatial contrast and solving the resulting optimization problem by an MM algorithm is considered in \citet[Sect. 3.2]{zhou2010graphics}.
Here $D$ is the finite difference matrix on the pixel grid.
To promote sharper contrast, we employ the anisotropy total variation (TV) penalty \citep{rudin1992nonlinear} and minimize
$-L(\lambda) + \rho \|D \lambda\|_1$.
Write $E = (e_{ij})$. Then
the PDHG algorithm (Sect. \ref{sec:pdhg}) can be applied. Put $K = [E^T,  D^T]^T$, $f(z, w) = \sum_{i} (- y_i \log z_i) + \rho \|w\|_1$, and $g(\lambda) = \mathbf{1}^T E \lambda + \delta_{+} (\lambda)$, where $\mathbf{1}$ is the all-one vector 
and $\delta_{+}$ is the $0/\infty$ indicator function for the nonnegative orthant.
Since $f(z, w)$ is separable in $z$ and $w$, applying iteration \eqref{eqn:cp} using the proximity operator \eqref{eqn:pet}, we obtain the following iteration:
\begin{align*}
z_i^{n+1} &= \textstyle\frac{1}{2}\Big[\big(z_i^n + \sigma (E\bar{\lambda}^{n})_i\big) - \big[\big(z_i^n + \sigma (E\bar{\lambda}^{n})_i\big)^2 + 4 \sigma y_i\big]^{1/2}\Big], \quad i=1, \dotsc, d \\
w^{n+1} &= P_{[-\rho, \rho]}(w^n + \sigma D \bar{\lambda}^{n})\\
\lambda^{n+1} &= P_+ (\lambda^n - \tau (E^T z^{n+1} + D^T w^{n+1} + E^T \mathbf{1})) \\
\bar{\lambda}^{n+1} &= 2 \lambda^{n+1} - \lambda^n,
\end{align*}
where $P_{[-\rho, \rho]}$ is elementwise projection to the interval $[-\rho, \rho]$.
Convergence is guaranteed if $\sigma \tau < 1/\| [E^T \; D^T] \|_2^2$.  
Scalability experiments were carried out with large Roland-Varadhan-Frangakis phantoms \citep{roland2007squared} using grid sizes $p = 300 \times 300$, $400 \times 400$, and $900 \times 900$, with number of detector pairs $d = 179,700$. 
Timing per 1000 iterations is reported in Table \ref{tab:pet}.
Both matrices $E$ and $D$ were distributed along the columns.
For reference, \citet{zhou2010graphics} use a $64 \times 64$ grid with $d = 2016$.
The total elapsed time decreases with more GPUs or nodes. 
The multi-GPU node could not run the $p = 810,000$ dataset, however, since the data size was too big to fit in the GPU memory.
Figure \ref{fig:pet_l1_xcat} illustrates TV reconstructions of a $p = 128 \times 128$ extended cardiac-torso (XCAT) phantom with $d=8128$ \citep{lim2018pet,ryu2018splitting}.
Results by a stochastic version of PDHG \citep{chambolle2018} are also provided. Each reconstruction was run for 20,000 iterations, which were sufficient for both algorithms to reach similar objective values. Those iterations took 20 to 35 seconds on a single GPU.

\paragraph{Multi-dimensional scaling}
The version of MDS considered in \citet[Sect. 3.3]{zhou2010graphics} minimizes the stress function
$f(\theta) = \sum_{i =1}^q \sum_{j \ne i} w_{ij}(y_{ij} - \|\theta_i - \theta_j \|_2)^2$
to map dissimilarity measures $y_{ij}$ between data point pairs $(i, j)$ to points $\theta = \left(\theta_1, \dots, \theta_q \right)^T$ in an Euclidean space of low dimension $p$,
where the $w_{ij}$ are the weights. \citet{zhou2010graphics} derive a parallel MM iteration
\begin{equation*}
\theta_{ik}^{n+1} = 
\textstyle
\left(\sum_{j \ne i} \left[\frac{y_{ij}}{\|\theta_{i}^n - \theta_{j}^n\|_2}(\theta_{ik}^n - \theta_{jk}^n) + (\theta_{ik}^n + \theta_{jk}^n)\right]\right) \big/ \left(2 \sum_{j \ne i}^m w_{ij} \right)
\end{equation*}
for $i = 1, \dots, q$ and $k = 1, \dots, p$.
We generated a [10,000] $\times$ 10,000 and a [100,000] $\times$ 100,000 pairwise dissimilarity matrices from samples of the 1,000-dimensional standard normal distribution. For reference, the dimension of the dissimilarity matrix used in \citet{zhou2010graphics} is $401 \times 401$. 
Elapsed time is reported in Table \ref{tab:mds1}.  
For $p=20$, the eight-GPU setting achieved a 5.32x-speedup compared to the single 36-core CPU AWS instance and a 6.13x-speedup compared to single GPU.
The larger experiment involved storing a distance matrix of size [100,000] $\times$ 100,000, which took 74.5 GB of memory. 
The multi-GPU node did not scale to run this experiment due to the memory limit.
On the other hand, we observed a 3.78x-speedup with 20 instances (720 cores) with respect to four instances (144 cores) of CPU nodes.

Appendix \ref{sec:numextra} contains further details on the experiments of this subsection.


\begin{table}
\caption{Runtime (in seconds) of NMF on simulated data for different inner dimensions $r$. ``$\times$'' denotes that the experiment could not run with a single data load to the device.}
\centering
\begin{tabular}{@{\extracolsep{10pt}}rrrrrrr@{}}
\hline 
configuration   & \multicolumn{3}{c}{10,000 $\times$ 10,000}        & \multicolumn{3}{c}{200,000 $\times$ 200,000}\\
& \multicolumn{3}{c}{10,000 iterations}        & \multicolumn{3}{c}{1000 iterations}
\\\cmidrule{2-4} \cmidrule{5-7}
   & $r=20$                 & $r=40$ & $r=60$ & $r=20$                 & $r=40$ & $r = 60$ \\ \hline
\multicolumn{7}{l}{GPUs}\\ \hline
1 & 164 & 168 & 174  & $\times$ & $\times$ & $\times$ \\
2 & 97 & 106 & 113  & $\times$ & $\times$ & $\times$ \\
4 & 66 & 78 & 90  & $\times$ & $\times$ & $\times$ \\
8 & 57 & 77 & 92  & $\times$ & $\times$ & $\times$ \\ \hline
\multicolumn{7}{l}{AWS EC2 \texttt{c5.18xlarge} instances} \\ \hline
4  & 205 & 310 & 430    & 1493                   & 1908   & 2232     \\
5  & 230 & 340 & 481    & 1326                   & 1652   & 2070     \\
8  & 328 & 390 & 536     & 937                    & 1044   & 1587     \\
10 & 420 & 559 & 643     & 737                    & 937    & 1179     \\
20 & 391 & 1094 & 1293     & 693                    & 818    & 1041    \\ \hline 
\end{tabular}\label{tab:nmf3}
\vspace{0.5cm}
\caption{Runtime (in seconds) comparison of 1000 iterations of TV-penalized PET.  We exploited sparse structures of $E$ and $D$. The number of detector pairs $d$ was fixed at 179,700.}\label{tab:pet}
\begin{tabular}{@{\extracolsep{10pt}}rrrr@{}}
\hline 
configuration   & $p=90,000$ & $p = 160,000$ & $p = 810,000$     \\ \hline
\multicolumn{4}{l}{GPUs}\\ \hline
1 & $\times$ & $\times$ & $\times$  \\
2 & 21 & 35 & $\times$  \\
4 & 19 & 31 & $\times$ \\
8 & 18 & 28 & $\times$ \\ \hline
\multicolumn{4}{l}{AWS EC2 \texttt{c5.18xlarge} instances} \\ \hline
4  & 36 & 49   & 210        \\
5  & 36 & 45   & 188       \\
8  & 33 & 39  & 178       \\
10 & 38 & 37   & 153         \\
20 & 26 & 28   & 131      \\ \hline 
\end{tabular}
\vspace{0.5cm}
\caption{Runtimes (in seconds) of 1000 iterations for MDS for different mapped dimensions $q$.}\label{tab:mds1}
\centering
\begin{tabular}{@{\extracolsep{10pt}}rrrrrrr@{}}
\hline 
configuration   & \multicolumn{3}{c}{10,000 datapoints} & \multicolumn{3}{c}{100,000 datapoints}      \\
& \multicolumn{3}{c}{10,000 iterations} & \multicolumn{3}{c}{1000 iterations}    
\\\cmidrule{2-4} \cmidrule{5-7} 
   & $q=20$                 & $q=40$ & $q=60$ & $q=20$                 & $q=40$ & $q=60$  \\ \hline
\multicolumn{7}{l}{GPUs}\\ \hline
1 & 368 & 376 & 384 & $\times$ & $\times$ & $\times$  \\
2 & 185 & 190 & 195 & $\times$ & $\times$ & $\times$  \\
4 & 100 & 103 & 108 & $\times$ & $\times$ & $\times$ \\
8 & 60 & 67 & 73 & $\times$ & $\times$ & $\times$ \\ \hline
\multicolumn{7}{l}{AWS EC2 \texttt{c5.18xlarge} instances} \\ \hline
4  & 424 & 568 & 596 & 3103                    & 3470    & 3296        \\
5  & 364 & 406 & 547 & 2634                    & 2700    & 2730       \\
8  & 350 & 425 & 520 & 1580                    & 1794    & 1834       \\
10 & 275 & 414 & 457 & 1490                    & 1454    & 1558         \\
20 & 319 & 440 & 511 & 820                    & 958    & 1043      \\ \hline 
\end{tabular}
\end{table}

\subsection{$\ell_1$-regularized Cox proportional hazards regression}\label{sec:cox}
We apply the proximal gradient method to $\ell_1$-regularized Cox proportional hazards regression \citep{cox1972regression}.
In this problem, we are given a covariate matrix $X \in \mathbb{R}^{m \times p}$, 
time-to-event $(t_1, \dotsc, t_m)$, and right-censoring time $(c_1, \dotsc, c_m)$ for individual $i=1,\dotsc,m$ as data.
The ``response'' 
is defined by  $y_i = \min \{t_i, c_i\}$ 
for each indivuali $i$,
and whether this individual is censored is indicated by $\delta_i= I_{\{t_i \le c_i\}}$.
The log partial likelihood of the Cox model is then
\begin{align*}
L (\beta) = \sum_{i=1}^m \delta_i \Big[\beta^T x_i - \log \big(\textstyle\sum_{j: y_j \ge y_i} \exp(\beta^T x_j)\big)\Big]. 
\end{align*}
Coordinate-descent-type approaches to this 
model
are proposed by \citet{suchard2013massive} and \citet{mittal2014high}.
    
To obtain a proximal gradient iteration, we need the gradient $\nabla L(\beta)$ and its Lipschitz constant. The gradient of the log partial likelihood is
\begin{align*}
\nabla L(\beta) = X^T (I-P) \delta,
\quad
\delta = (\delta_1, \dotsc, \delta_m)^T,
\end{align*} 
where we define the matrix $P = (\pi_{ij})$ with
$\pi_{ij} = I(y_i \ge y_j) w_i/W_j$;
$w_i = \exp(x_i^T \beta)$, $W_j = \sum_{i: y_i \ge y_j} w_i$.
A Lipschitz constant of $\nabla L(\beta)$ can be found by finding an upper bound of 
the Hessian of $L(\beta)$:
$$
\nabla^2 L(\beta) = X^T (P \mathrm{diag}(\delta) P^T - \mathrm{diag}(P \delta))X. 
$$
Note $\|P\|_2 \le 1$, since the sum of each row of $P$ is 1. It follows that $\|\nabla^2 L(\beta)\|_2 \le 2 \|X\|_2^2$, 
and $\|X\|_2$ can be quickly computed by using the power iteration \citep{golub2013matrix}.

We introduce an $\ell_1$-penalty to the log partial likelihood in order to enforce sparsity in the regression coefficients
and use the proximal gradient descent to estimate $\beta$ by putting $g(\beta) = -L(\beta)$, $f(\beta) = \lambda \|\beta\|_1$. 
Then the 
iteration is:
\begin{align*}
w_i^{n+1} &= \exp(x_i^T \beta); \;\; W_j^{n+1} = \textstyle\sum_{i: y_i \ge y_j} w_i^{n+1}\\
\pi_{ij}^{n+1} &= I_{\{t_i \ge t_j\}} w_i^{n+1} / W_j^{n+1} \\
\Delta^{n+1} &= X^T (I - P^{n+1}) \delta, \;\; \text{where $P^{n+1} = (\pi_{ij}^{n+1})$} \\
\beta^{n+1} &= \mathcal{S}_{\lambda}(\beta^n + \sigma \Delta^{n+1}).
\end{align*}

If the data are sorted in 
descending
order of $y_i$, the $W_j^n$ can be computed by cumulative summing $(w_1,\dotsc, w_m)$ in the proper order. 
A CUDA kernel 
for this operation is readily available in PyTorch. 
The soft-thresholding operator $\mathcal{S}_\lambda (x)$ is also implemented in PyTorch.
We can write a simple proximal gradient descent routine 
for the Cox regression as in Listing \ref{code:cox}, assuming no ties in $y_i$'s. 

\begin{lstfloat}[ht!]
\begin{lstlisting}[basicstyle=\scriptsize\ttfamily,label=code:cox, caption=PyTorch code for the proximal gradient method for $\ell_1$-regularized Cox regression.]
import torch
import torch.distributed as dist
from dist_stat import distmat
from dist_stat.distmat import distgen_uniform, distgen_normal
dist.init_process_group('mpi')
rank = dist.get_rank()
size = dist.get_world_size()
device = 'cuda:{}'.format(rank) # 'cpu' for CPU computing
if device.startswith('cuda'): torch.cuda.set_device(rank)
n = 10000
p = 10000
max_iter = 10000
TType = torch.cuda.FloatTensor if device.startswith('cuda') else torch.DoubleTensor
X = distgen_normal(p, n, TType=TType).t() # [p] x n transposed => n x [p].
delta = torch.multinomial(torch.tensor([1., 1.]), n, replacement=True).float().view(-1, 1).type(TType) # censoring indicator, n x 1, Bernoulli(0.5). 
delta_dist = distmat.dist_data(delta, TType=TType) # distribute delta to create [n] x 1 data
beta = distmat.dist_data(torch.zeros((p, 1)).type(TType), TType=TType) #[p] x 1
Xt = X.t() # transpose. [p] x n 
sigma = 0.00001 # step size
lambd = 0.00001 # penalty parameter
soft_threshold = torch.nn.Softshrink(lambd) # soft-thresholding

# Data points are assumed to be sorted in decreasing order of observed time.
y_local = torch.arange(n, 0, step=-1).view(-1, 1).type(TType) # local n x 1
y_dist = distmat.dist_data(y_local, TType=TType) # distributed [n] x 1
pi_ind = (y_dist - y_local.t() >= 0).type(TType)

Xbeta = distmat.mm(X, beta) # Scenario 5 (default for n x [p] times [p] x 1)

for i in range(max_iter):
    w = Xbeta.exp() # n x 1
    W = w.cumsum(0) # n x 1
    dist.barrier() # wait until the distributed computation above is finished

    w_dist = distmat.dist_data(w, TType=TType) # distribute w. [n] x 1.

    pi = (w_dist / W.t()) * pi_ind # [n] x n.
    pd = distmat.mm(pi, delta) # Scenario 4.
    dmpd = delta_dist - pd # [n] x 1.
    grad = distmat.mm(Xt, dmpd) # Scenario 1.
    beta = (beta + grad * sigma).apply(soft_threshold) # [p] x 1
    Xbeta = distmat.mm(X, beta) # Scenario 5.

    expXbeta = (Xbeta).exp() # n x 1
    obj = distmat.mm(delta.t(), (Xbeta - (expXbeta.cumsum(0)).log())) \
        - lambd * beta.abs().sum() # mm: local computation.  
    print(i, obj.item())
\end{lstlisting}
\end{lstfloat}

A synthetic
data matrix $X \in \mathbb{R}^{m \times [p]}$, distributed along the columns, was sampled from the standard normal distribution.
The algorithm was designed to keep a copy of the estimand $\beta$ in every device.
All the numerical experiments were carried out with double precision even for GPUs, for the following reason. 
For a very small value of $\lambda$ (we used $\lambda = 10^{-5}$), when single precision was used in GPUs, the estimate quickly tended to ``not a number (NaN)''s due to numerical instability of the CUDA kernel. 
Double-precision did not generate such a problem. 
Although desktop GPU models such as Nvidia GTX and Titan X are not optimized for double precision floating-point operations and is known to be 32 times slower for double precision operations than single precision operations,  
this does not necessarily mean that the total computation time is 32 times slower, since latency takes a significant portion of the total computation time in GPU computing.


In order to demonstrate the scalability of our approach, elapsed times for 10,000 $\times$ [10,000] and 100,000 $\times$ [200,000] simulated data are reported in Table \ref{tab:cox3}. We can see 3.92x speedup from 4 nodes to 20 nodes in the virtual cluster.
Even with double-precision arithmetics, eight GPUs could achieve a 6.30x-speedup over the single 36-core CPU instance. 
As expected,
virtual clusters in a cloud exhibited better scalability.

\begin{table}[]
\caption{Runtime comparison of $\ell_1$-regularized Cox regression over multi-node virtual cluster on AWS EC2. Elapsed time (in seconds) after 1000 iterations.}\label{tab:cox3}
\begin{tabular}{@{\extracolsep{10pt}}rrr@{}}
\hline 
configuration   & $10,000 \times [10,000]$ & $100,000 \times [200,000]$      \\
& 10,000 iterations & 1,000 iterations    \\ \hline
\multicolumn{3}{l}{GPUs}\\ \hline
1 & 386 & $\times$  \\
2 & 204 & $\times$  \\
4 & 123 & $\times$ \\
8 & 92 & $\times$ \\ \hline
\multicolumn{3}{l}{AWS EC2 \texttt{c5.18xlarge} instances} \\ \hline
1 & 580 & $\times$ \\
2 & 309  & $\times$      \\
4  & 217    & 1507        \\
5  & 170   & 1535       \\
8  & 145   & 775       \\
10 & 132    & 617         \\
20 & 148    & 384      \\ \hline 
\end{tabular}
\end{table}


\subsection{Genome-wide survival analysis of the UK Biobank dataset}\label{sec:biobank}
We demonstrate a real-world application of $\ell_1$-regularized Cox proportional hazards regression to genome-wide survival analysis for Type 2 Diabetes (T2D).
We used a UK Biobank dataset \citep{ukbiobank2015} that contains information on approximately 800,000 single nucleotide polymorphisms (SNPs) of 500,000 individual subjects recruited from the United Kingdom.
After filtering SNPs for quality control and subjects for the exclusion of Type 1 Diabetes patients, 402,297 subjects including 17,994 T2D patients and 470,189 SNPs remained. We randomly sampled 200,000 subjects including 8,995 T2D patients for our analysis. Any missing genotype was imputed with the column mean.
Along with the SNPs, sex and top ten principal components were included as unpenalized covariates to adjust for population-specific variations. The resulting dataset was 701 GB with double-precision.

The analysis for this large-scale genome-wide dataset was conducted as follows.
Incidence of T2D was used as the event ($\delta_i = 1$) and the age of onset was used as survival time $y_i$. For non-T2D subjects ($\delta_i = 0$), age at the last visit was used as $y_i$. 
We chose 63 different values of the regularization parameter $\lambda$ in the range $[0.7 \times 10^{-9}, 1.6 \times 10^{-8}]$, with which 0 to 111 SNPs were selected.
For each value of $\lambda$, the $\ell_1$-regularized Cox regression model of Section \ref{sec:cox} was fitted. Every run converged after at most 2080 iterations that took less than 2800 seconds using 20 \texttt{c5.18xlarge} instances from AWS EC2. 

The SNPs were ranked based on the largest value of $\lambda$ with which a SNP is selected.
(No variables were removed once selected within the range of $\lambda$ used. The regularization path and the full list of the selected SNPs are available in Appendix \ref{sec:snps}.)
Among the 111 SNPs selected, three of the top four selections were located on TCF7L2, whose association with T2D is well-known \citep{scott2007genome,wellcome2007genome}. 
Also prominently selected were SNPs from genes SLC45A2 and HERC2, whose variants are known to be associated with skin, eye, and hair pigmentation \citep{cook2009analysis}. This is possibly due to the dominantly European population in the UK Biobank study. 
Mapped genes for 24 SNPs out of the selected 111 were also reported in \citet{mahajan2018fine}, a meta-analysis of 32 genome-wide association studies (GWAS) for about 898,130 individuals of European ancestry; see Tables \ref{tab:snps_pt1} and \ref{tab:snps_pt2} for details.
We then conducted an unpenalized Cox regression analysis using the 111 selected SNPs. The nine SNPs with $p$-values less than 0.01 are listed in Table \ref{tab:snps_pv}.
The locations in Table \ref{tab:snps_pv} are with respect to the reference genome GRCh37 \citep{church2011modernizing}, and mapped genes were predicted by the Ensembl Variant Effect Predictor \citep{mclaren2016ensembl}.
Among these nine SNPs, three of them were directly shown to be associated with T2D (\citet{wellcome2007genome} and \citet{dupuis2010new} for rs4506565, \citet{voight2010twelve} for rs8042680, \citet{ng2014meta} for rs343092). Three other SNPs have mapped genes reported to be associated with T2D in \citet{mahajan2018fine}: rs12243326 on TCF7L2, rs343092 on HMGA2, and rs231354 on KCNQ1.


Although the interpretation of the results requires additional sub-analysis, the result shows the promise of joint association analysis using multiple regression models. In GWAS it is customary to analyze the data on SNP-by-SNP basis. Among the mapped genes harboring the 111 SNPs selected by our half-million-variate regression analysis are CPLX3 and CACNA1A, associated with regulation of insulin secretion, and SEMA7A and HLA-DRA involved with inflammatory responses (based on DAVID \citep{david2,david1}). These genes might have been missed in conventional univariate analysis of T2D due to 
nominally
moderate statistical significance. 
Joint GWAS may overcome such a limitation and is possible by combining the computing power of modern HPC and scalable algorithms.

\begin{table}[]
\resizebox{\textwidth}{!}{
\begin{threeparttable}
\caption{SNPs with $p$-values of less than 0.01 on unpenalized Cox regression with  variables selected by $\ell_1$-penalized Cox regression}\label{tab:snps_pv}
\begin{tabular}{lrrllrlrr}
\hline
SNP ID     & Chr.     & Location  & A1\tnote{A} & A2\tnote{B} & MAF\tnote{C} &  Mapped Gene & Coefficient & $p$-value                    \\ \hline
rs4506565  & 10         & 114756041 & A       & \textbf{T}     & 0.238   & TCF7L2  & 2.810e-1 & $<$2e-16                \\
rs12243326 & 10         & 114788815 & \textbf{C}       & T     & 0.249   & TCF7L2  & 1.963e-1 & 0.003467              \\
rs8042680  & 15         & 91521337  & \textbf{A}       & C     & 0.277   & PRC1  & 2.667e-1 & 0.005052 \\
rs343092   & 12         & 66250940  & \textbf{T}       & G     & 0.463 & HMGA2  & $-$7.204e-2 & 0.000400 \\
rs7899137  & 10         & 76668462  & \textbf{A}       & C     & 0.289   & KAT6B  & $-$4.776e-2 & 0.002166 \\
rs8180897  & 8          & 121699907 & A       & \textbf{G}     & 0.445   & SNTB1  & 6.361e-2  & 0.000149 \\
rs10416717 & 19         & 13521528  & A       & \textbf{G}     & 0.470   & CACNA1A  & 5.965e-2 & 0.009474 \\
rs231354   & 11         & 2706351   & \textbf{C}       & T     & 0.329   & KCNQ1  & 4.861e-2 & 0.001604 \\
rs9268644  & 6          & 32408044  & \textbf{C}       & A     & 0.282   & HLA-DRA  & 6.589e-2 & 2.11e-5 \\ \hline
\end{tabular}
\begin{tablenotes}
\item[A] Minor allele, \item[B] Major allele, \item[C] Minor allele frequency. The boldface indicates the risk allele determined by the reference allele and the sign of the regression coefficient. 
\end{tablenotes}
\end{threeparttable}
}
\end{table}

\section{Discussion}\label{sec:discuss}

Abstractions of highly complex computing operations have rapidly evolved over the last decade.
In this article, we have explained how statisticians can benefit from this evolution.
We have seen how deep learning technology is relevant to high-performance statistical computing. 
We have also demonstrated that 
many useful tools for incorporating accelerators and computing clusters  have been created.
Unfortunately, such developments have been mainly made in languages other than R, particularly in Python, with which statisticians may not be familiar with.
Although there are libraries that deal with simple parallel computation in R, there are common issues with these libraries. First, the libraries do not easily incorporate GPUs that might significantly speed up computation. Second, it is hard to write more full-fledged parallel programs without directly writing code in C or C++. 
This two-language problem calls for statisticians to take a second look at Python.
Fortunately, this language is not hard to learn, and younger generations are quite familiar with it. 
A remedy from the R side may be either developing more user-friendly interfaces for the distributed-memory environment, with help from those who are engaged in computer engineering, or 
writing a good wrapper for the important Python libraries.
A Python interface to R may be a good starting point.
For example, 
R package \texttt{reticulate} \citep{reticulate} is a basis of
other
interfaces 
packages
to PyTorch \citep[\texttt{rTorch}, ][]{rtorch} and TensorFlow \citep[also called \texttt{tensorflow},][]{rtensorflow}.

By making use of multiple CPU nodes or a multi-GPU workstation, the methods discussed in the current article can be applied efficiently even when the dataset exceeds several tens of gigabytes.
The advantages of 
engaging multiple compute devices
are two-fold. First, we can take advantage of data parallelism with more computing cores, accelerating the computation. Second, we can push the limit of the size of the dataset to analyze. 
As cloud providers now support virtual clusters better suited for HPC, statisticians can deal with bigger problems utilizing such services, using up to several thousand cores easily.  
When the data do not fit into the GPU memory (e.g., the UK Biobank example), it is still possible to carry out computation by moving partitions of the data in and out of GPUs. However, this is impractical because of slow communication between the main and GPU memories. On the other hand, virtual clusters are scalable with this size of data.

Loss of accuracy due to the default single precision of GPU arithmetic, prominent in our proportional hazards regression example, can be solved by purchasing scientifically-oriented GPUs with better double precision supports. 
Another option is 
migrating to the cloud: 
for example, the \texttt{P2} and \texttt{P3} instances in AWS support scientific GPUs. 
Nevertheless, desktop GPUs with double precision arithmetic turned on could achieve more than 10-fold speedup over CPU,
even though  double precision floating-point operations are 32 times slower than single precision.

Most of the highly parallelizable algorithms considered in Section \ref{sec:algo} require no more than the first-order derivative information, and this feature contributes to their low per-iteration complexity and parallelizability. 
As mentioned in Section \ref{sec:intro}, some second-order methods for sparse regression \citep{li2018highly,huang2018constructive,jin2019unified} maintain the set of active variables (of nonzero coefficients), and only these are involved in the Newton-Raphson step. Thus if the solution is sparse, the cost of solving the relevant linear system is moderate. With distributed matrix computation exemplified with \texttt{distmat}, residual and gradients can be computed in a  distributed fashion and the linear system can be solved after gathering active variables into the master device.


A major weakness of the present approach is that its effectiveness can be degraded by the communication cost between the nodes and devices. 
One way to avoid this issue is by using high-speed interconnection between the nodes and devices. In multi-CPU clusters, this can be realized by a high-speed interconnection technology such as InfiniBand. Even when such an environment is not affordable, we may still use relatively high-speed connection equipped with instances from a cloud. The network bandwidth of 25Gbps supported for \texttt{c5.18xlarge} instances of AWS was quite effective in our experiments. 
Reducing the number of communication rounds and iterations with theoretical guarantees, 
for example, by one-shot averaging \citep{zhang2013communication,duchi2014optimality,lee2015communication}, by using global first-order information and local higher-order information \citep{wang2017efficient,jordan2018communication,fan2019communication},
or by quantization \citep{pmlr-v97-tang19d,pmlr-v108-liu20a},
is an active area of current research.

Although PyTorch has been advocated throughout this article, it is not the only path towards easy-to-use programming models in shared- and distributed-memory programming environments. 
A possible alternative is Julia \citep{bezanson2017julia}, 
in which
data can reside in a wide variety of environments, such as GPUs \citep{besard2018effective} and multiple CPU nodes implementing the distributed memory model \citep{distributedarrays,mpiarrays}.  
While its long-term support release of version 1.0.5 in September 2019 is still fresh, Julia has the potential to be a powerful tool for statistical HPC
once the platforms and user community 
mature. 
\section*{Acknowledgements}
This research was partially funded by the National Research Foundation of Korea (NRF) grant funded by the Korea government (MSIT)
(2019R1A2C1007126, JHW; \linebreak 2020R1A6A3A03037675, SK), the Collaboratory Fellowship program of the UCLA Institute for Quantitative
\& Computational Bioscience (SK), AWS Cloud Credit for Research (SK and JHW), and grants from National Institutes of Health (R35GM141798, HZ; R01HG006139, HZ and JJZ; K01DK106116, JJZ; R21HL150374, JJZ) and National Science Foundation (DMS-2054253, HZ and JJZ).
\bibliography{dist_mpicuda.bib}
\appendix
\fontsize{10}{11}\selectfont
\counterwithin{figure}{section}
\counterwithin{table}{section}
\counterwithin{lstlisting}{section}
\newpage
\section{A brief introduction to PyTorch}\label{sec:pytorch}

In this section, we introduce simple operations on PyTorch. Note that Python uses 0-based, row-major ordering, like C and C++ (R is 1-based, column-major ordering). 
First we import the PyTorch library. This is equvalent to \texttt{library()} in R.
{
\begin{Shaded}
\begin{Highlighting}[fontsize=\footnotesize]
\ImportTok{import} \NormalTok{torch}
\end{Highlighting}
\end{Shaded}
}
\subsection*{Tensor creation}
The following is equivalent to \texttt{set.seed()} in R.
{
\begin{Shaded}
\begin{Highlighting}[fontsize=\footnotesize]
\NormalTok{torch.manual_seed(}\DecValTok{100}\NormalTok{) }
\end{Highlighting}
\end{Shaded}
}
One may create an uninitialized tensor. This creates a $3 \times 4$ tensor (matrix).
{
\begin{Shaded}
\begin{Highlighting}[fontsize=\footnotesize]
\NormalTok{torch.empty(}\DecValTok{3}\NormalTok{, }\DecValTok{4}\NormalTok{) }\CommentTok{# uninitialized tensor}
\end{Highlighting}
\end{Shaded}

\begin{Verbatim}[fontsize=\footnotesize]
tensor([[-393462160144990208.0000,                   0.0000,
         -393462160144990208.0000,                   0.0000],
        [                  0.0000,                   0.0000,
                           0.0000,                   0.0000],
        [                  0.0000,                   0.0000,
                           0.0000,                   0.0000]])
\end{Verbatim}
}
This generates a tensor initialized with random values from $(0,1)$.
{
\begin{Shaded}
\begin{Highlighting}[fontsize=\footnotesize]
\NormalTok{y }\OperatorTok{=} \NormalTok{torch.rand(}\DecValTok{3}\NormalTok{, }\DecValTok{4}\NormalTok{) }\CommentTok{# from Unif(0, 1)}
\end{Highlighting}
\end{Shaded}

\begin{Verbatim}[fontsize=\footnotesize]
tensor([[0.1117, 0.8158, 0.2626, 0.4839],
        [0.6765, 0.7539, 0.2627, 0.0428],
        [0.2080, 0.1180, 0.1217, 0.7356]])
\end{Verbatim}
}
We can also generate a tensor filled with zeros or ones.
{
\begin{Shaded}
\begin{Highlighting}[fontsize=\footnotesize]
\NormalTok{z }\OperatorTok{=} \NormalTok{torch.ones(}\DecValTok{3}\NormalTok{, }\DecValTok{4}\NormalTok{)}\CommentTok{ # torch.zeros(3, 4)}
\end{Highlighting}
\end{Shaded}

\begin{Verbatim}[fontsize=\footnotesize]
tensor([[1., 1., 1., 1.],
        [1., 1., 1., 1.],
        [1., 1., 1., 1.]])
\end{Verbatim}
}
A tensor can be created from standard Python data.
{
\begin{Shaded}
\begin{Highlighting}[fontsize=\footnotesize]
\NormalTok{w }\OperatorTok{=} \NormalTok{torch.tensor([}\DecValTok{3}\NormalTok{, }\DecValTok{4}\NormalTok{, }\DecValTok{5}\NormalTok{, }\DecValTok{6}\NormalTok{])}
\end{Highlighting}
\end{Shaded}

\begin{Verbatim}[fontsize=\footnotesize]
tensor([3, 4, 5, 6])
\end{Verbatim}
}
\subsection*{Indexing}
The following is the standard method of indexing tensors.
{
\begin{Shaded}
\begin{Highlighting}[fontsize=\footnotesize]
\NormalTok{y[}\DecValTok{2}\NormalTok{, }\DecValTok{3}\NormalTok{] }\CommentTok{# indexing: zero-based, returns a 0-dimensional tensor}
\end{Highlighting}
\end{Shaded}

\begin{Verbatim}[fontsize=\footnotesize]
tensor(0.7356)
\end{Verbatim}
}
The indexing always returns a (sub)tensor, even for scalars (treated as zero-dimensional tensors). A standard Python number can be returned by using \texttt{.item()}.  
{
\begin{Shaded}
\begin{Highlighting}[fontsize=\footnotesize]
\NormalTok{y[}\DecValTok{2}\NormalTok{, }\DecValTok{3}\NormalTok{].item() }\CommentTok{# A standard Python floating-point number}
\end{Highlighting}
\end{Shaded}

\begin{Verbatim}[fontsize=\footnotesize]
0.7355988621711731
\end{Verbatim}
}
To get a column from a tensor, we use the indexing as below. The syntax is similar but slightly different from R.
{
\begin{Shaded}
\begin{Highlighting}[fontsize=\footnotesize]
\NormalTok{y[:, }\DecValTok{3}\NormalTok{] }\CommentTok{# 3rd column. The leftmost column is 0th. cf. y[, 4] in R}
\end{Highlighting}
\end{Shaded}

\begin{Verbatim}[fontsize=\footnotesize]
tensor([0.4839, 0.0428, 0.7356])
\end{Verbatim}
}
The following is for taking a row.
{
\begin{Shaded}
\begin{Highlighting}[fontsize=\footnotesize]
\NormalTok{y[}\DecValTok{2}\NormalTok{, :] }\CommentTok{# 2nd row. The top row is 0th. cf. y[3, ] in R}
\end{Highlighting}
\end{Shaded}

\begin{Verbatim}[fontsize=\footnotesize]
tensor([0.2080, 0.1180, 0.1217, 0.7356])
\end{Verbatim}
}
\subsection*{Simple operations}
Here we provide an example of simple operations on PyTorch. Addition using the operator `+' acts just like anyone can expect:
{
\begin{Shaded}
\begin{Highlighting}[fontsize=\footnotesize]
\NormalTok{x }\OperatorTok{=} \NormalTok{y }\OperatorTok{+} \NormalTok{z }\CommentTok{# a simple addition.}
\end{Highlighting}
\end{Shaded}

\begin{Verbatim}[fontsize=\footnotesize]
tensor([[1.1117, 1.8158, 1.2626, 1.4839],
        [1.6765, 1.7539, 1.2627, 1.0428],
        [1.2080, 1.1180, 1.1217, 1.7356]])
\end{Verbatim}
}
Here is another form of addition.
{
\begin{Shaded}
\begin{Highlighting}[fontsize=\footnotesize]
\NormalTok{x }\OperatorTok{=} \NormalTok{torch.add(y, z) }\CommentTok{# another syntax for addition}
\end{Highlighting}
\end{Shaded}
}

The operators ending with an underscore (\_) changes the value of the tensor in-place.
{
\begin{Shaded}
\begin{Highlighting}[fontsize=\footnotesize]
\NormalTok{y.add_(z) }\CommentTok{# in-place addition}
\end{Highlighting}
\end{Shaded}

\begin{Verbatim}[fontsize=\footnotesize]
tensor([[1.1117, 1.8158, 1.2626, 1.4839],
        [1.6765, 1.7539, 1.2627, 1.0428],
        [1.2080, 1.1180, 1.1217, 1.7356]])
\end{Verbatim}
}
\subsection*{Concatenation}
We can concatenate the tensors using the function \texttt{cat()}, which resembles \texttt{c()}, \texttt{cbind()}, and \texttt{rbind()} in R. The second argument indicates the dimension that the tensors are concatenated along: 
zero signifies concatenation by rows while one by columns.
{
\begin{Shaded}
\begin{Highlighting}[fontsize=\footnotesize]
\NormalTok{torch.cat((y, z), }\DecValTok{0}\NormalTok{) }\CommentTok{# along the rows}
\end{Highlighting}
\end{Shaded}

\begin{Verbatim}[fontsize=\footnotesize]
tensor([[1.1117, 1.8158, 1.2626, 1.4839],
        [1.6765, 1.7539, 1.2627, 1.0428],
        [1.2080, 1.1180, 1.1217, 1.7356],
        [1.0000, 1.0000, 1.0000, 1.0000],
        [1.0000, 1.0000, 1.0000, 1.0000],
        [1.0000, 1.0000, 1.0000, 1.0000]])
\end{Verbatim}
}
{
\begin{Shaded}
\begin{Highlighting}[fontsize=\footnotesize]
\NormalTok{torch.cat((y, z), }\DecValTok{1}\NormalTok{) }\CommentTok{# along the columns}
\end{Highlighting}
\end{Shaded}

\begin{Verbatim}[fontsize=\scriptsize]
tensor([[1.1117, 1.8158, 1.2626, 1.4839, 1.0000, 1.0000, 1.0000, 1.0000],
        [1.6765, 1.7539, 1.2627, 1.0428, 1.0000, 1.0000, 1.0000, 1.0000],
        [1.2080, 1.1180, 1.1217, 1.7356, 1.0000, 1.0000, 1.0000, 1.0000]])
\end{Verbatim}
}

\subsection*{Reshaping}
One can reshape a tensor, like changing the attribute \texttt{dim} in R.
{
\begin{Shaded}
\begin{Highlighting}[fontsize=\footnotesize]
\NormalTok{y.view(}\DecValTok{12}\NormalTok{) }\CommentTok{# 1-dimensional array}
\end{Highlighting}
\end{Shaded}

\begin{Verbatim}[fontsize=\scriptsize]
tensor([1.1117, 1.8158, 1.2626, 1.4839, 1.6765, 1.7539, 1.2627, 1.0428, 1.2080,
        1.1180, 1.1217, 1.7356])
\end{Verbatim}
}
Up to one of the arguments of \texttt{view()} can be $-1$. The size of the reshaped tensor is inferred from the other dimensions.
{
\begin{Shaded}
\begin{Highlighting}[fontsize=\footnotesize]
\CommentTok{# reshape into (6)-by-2 tensor;}
\CommentTok{# (6) is inferred from the other dimension}
\NormalTok{y.view(}\OperatorTok{-}\DecValTok{1}\NormalTok{, }\DecValTok{2}\NormalTok{) }
\end{Highlighting}
\end{Shaded}

\begin{Verbatim}[fontsize=\footnotesize]
tensor([[1.1117, 1.8158],
        [1.2626, 1.4839],
        [1.6765, 1.7539],
        [1.2627, 1.0428],
        [1.2080, 1.1180],
        [1.1217, 1.7356]])
\end{Verbatim}
}
\section{AWS EC2 and ParallelCluster}\label{sec:cloud}

We used AWS Elastic Compute Cloud (EC2) via CfnCluster throughout our
multi CPU-node experiments, which is updated to ParallelCluster after we had completed the experiments. In this section, we instruct how to use
ParallelCluster via Amazon Web Services. This section is structured into
three parts: 
set up AWS account, configure, and run a job on ParallelCluster.
We refer the readers to the official documentation\footnote{\url{https://docs.aws.amazon.com/parallelcluster/index.html}} and an AWS whitepaper\footnote{\url{https://d1.awsstatic.com/Projects/P4114756/deploy-elastic-hpc-cluster_project.a12a8c61339522e21262da10a6b43a3678099220.pdf}} for further details.

\subsection{Overview}

A virtual cluster created by ParallelCluster consists of two types of
\emph{instances} in EC2: a \emph{master} instance and multiple \emph{worker}
instances. The master instance manages \emph{jobs} through a queue on a
\emph{job scheduler} and several AWS services such as \emph{Simple Queue
Service} and \emph{Auto Scaling Group}. When a virtual cluster is
created, a volume of \emph{Elastic Block Store (EBS)} is created and
automatically attached to the master instance. It is shared over the workers, forming a \emph{shared file system}.
The software necessary for the jobs are installed in this file system, and a
script to set up the environment variables for the tools is utilized. 
While the master instance does not directly take part in the
actual computation, the speed of network on the shared file system
depends on the instance type of the master instance. If the jobs depend
on the shared dataset, the master instance has to allow fast enough
network speed. The actual computation is performed on the worker
instances. Each worker has access to the shared file system where the
necessary tools and data reside. The network speed between workers
depends on the worker instance type.

\subsection{Glossary}

We briefly introduce some of the key concepts regarding the AWS and
cluster computing in this subsection.

Some of the basic concepts from AWS are shown below:

\begin{itemize}

\item
  Instance: a virtual computer on AWS EC2. 
  There are various types of instances, which are distinguished by the number of cores, memory size, network speed, etc.
  \texttt{c5.18xlarge} is prominently utilized in our experiments.\footnote{See
  \url{https://aws.amazon.com/en/ec2/instance-types/} for the full list
  of types of instances.} 
\item
  Region: 
  a region (e.g., North Virginia, Ohio, North California, Oregon, Hong. Kong, Seoul, Tokyo) is completely independent from the other regions, and the data transfer between regions are charged.
\item
  Availability zone: there are a handful of availability zones in each
  region. Each availability zone is isolated, but availability zones in the same region are interconnected with a low-latency network. Note that a virtual cluster created by ParallelCluster is tied to a single availability zone.
\end{itemize}

Listed below are some, but not all, of the AWS services involved in
ParallelCluster. They are all managed automatically through
ParallelCluster and can be modified through the AWS console.

\begin{itemize}

\item
  Elastic Compute Cloud (EC2): the core service of AWS that allows users
  to rent virtual computers. There are three methods of payment
  available:

  \begin{itemize}
  
  \item
    On-demand: hourly charged, without risk of interruption.
  \item
    Spot: bid-based charging. Serviced at up to 70\%-discounted rate,
    but is interrupted if the price goes higher than the bid price.
  \item
    Reserved: one-time payment at discounted rate.
  \end{itemize}
\item
  Elastic Block Store (EBS): persistent block storage volume for EC2
  instances, e.g.~a solid-state drive (SSD). In ParallelCluster, each instance is started
  with a root EBS volume exclusive to each instance.
\item
  CloudFormation: An interface that describes and provisions the cloud
  resources.
\item
  Simple Queue Service: the actual job queue is served through message
  passing between EC2 instances.
\item
  CloudWatch: monitors and manages the cloud.
\item
  Auto Scaling Group: a collection of EC2 instances with similar
  characteristics. The number of instances is automatically scaled based
  on criteria defined over CloudWatch.
\item
  Identity and Access Management (IAM): An IAM user is an ``entity that
  [one] creates in AWS to represent the person or application that uses it
  to interact with AWS.''\footnote{\url{https://docs.aws.amazon.com/IAM/latest/UserGuide/id_users.html}}
  Each IAM user is granted certain permissions determined by the root
  user. As there are many services involved in ParallelCluster, it is
  recommended to use an IAM user with full permission.
\item
  Virtual Private Cloud (VPC): a VPC is a dedicated virtual network
  exclusive to the user, isolated from any other VPCs, which spans all
  the availability zones in one region. A subnet is a subnetwork in VPC
  exclusive to a single availity zone.\footnote{\url{https://docs.aws.amazon.com/vpc/latest/userguide/VPC_Subnets.html}}
\item
  Security Group (SG): A security group acts as a ``virtural firewall
  that controls the traffic for one or more instances.''\footnote{\url{https://docs.aws.amazon.com/AWSEC2/latest/UserGuide/using-network-security.html}}
\end{itemize}

Here are some of the concepts related to cluster computing:

\begin{itemize}

\item
  Shared file system: for multiple instances to work on the same data,
  it is convenient to have a file system that can be accessed by all the
  instances involved. In ParallelCluster, it is implemented as an
  additional EBS volume attached to the master instance. All the worker
  instances can access this volume, and its speed of network depends on
  the instance type of the master instance.
\item
  Job: a unit of execution, defined by either a single command or a job
  script.
\item
  Queue: a data structure containing jobs to run. Jobs in a queue are
  managed and prioritized by a job scheduler.
\item
  Master: an instance that manages the job scheduler.
\item
  Worker: an instance that executes the jobs.
\item
  Job scheduler: an application program that controls the execution of jobs over a
  cluster. e.g.~Sun Grid Engine, Torque, Slurm, etc. The Sun Grid Engine
  (SGE) was used for our experiments.
\end{itemize}

Several SGE commands are as follows:

\begin{itemize}

\item
  \texttt{qsub}: submits a job to the job queue
\item
  \texttt{qdel}: removes a job on the job queue
\item
  \texttt{qstat}: shows the current status of the queue
\item
  \texttt{qhost}: shows the current list of workers
\end{itemize}

\subsection{Prerequisites}

The following are needed before we proceed. Most of these might be
considered the first steps to use AWS.

\begin{itemize}

\item
  Access keys with administrative privileges: Access keys are
  credentials for IAM users and root users. They consist of access key
  ID (analogous to username) and secret access key (analogous to
  passwords). They should be kept confidential. It is recommended to
  create a temporary IAM user with administrative privilage and create
  an access key ID and a secret access key for the IAM user. They can be
  created in the AWS console (or the IAM console for an IAM user).\footnote{\url{https://docs.aws.amazon.com/IAM/latest/UserGuide/id_credentials_access-keys.html}}
\item
  A VPC and a subnet: A VPC for each region and a subnet for each availability zone is
  created by default. One may use these default VPC and subnet or
  newly-created ones.
\item
  A security group: One may use a default security group or a
  newly-created one.
\item
  A key pair that allows the user to access the cloud via SSH: Amazon
  EC2 uses public-key cryptography for login credentials. Each EC2
  instance is configured with a public key, and the user has to access
  this instance using the matching private key. It can be generated and
  managed on AWS EC2 console as well as the user's terminal.\footnote{\url{https://docs.aws.amazon.com/AWSEC2/latest/UserGuide/ec2-key-pairs.html}}
\end{itemize}

\subsection{Installation}

First, we install the ParallelCluster command line interface (CLI) on a
local machine. ParallelCluster command line interface is distributed
through the standard Python Package Index (PyPI), so one may install it
through \texttt{pip}, the standard package-installing command for
Python. One may install ParallelCluster by executing the following on
the command line:

\begin{Shaded}
\begin{Highlighting}[fontsize=\footnotesize]
\FunctionTok{sudo}\NormalTok{ pip install aws-parallelcluster}
\end{Highlighting}
\end{Shaded}

\subsection{Configuration}

Once ParallelCluster is installed on a local machine, an initial
configuration is needed. It can be done by various ways, but the easiest
way is through the command below:

\begin{Shaded}
\begin{Highlighting}[fontsize=\footnotesize]
\ExtensionTok{pcluster}\NormalTok{ configure}
\end{Highlighting}
\end{Shaded}

Then, the interactive dialog to setup ParallelCluster appears:

\begin{Shaded}
\begin{Highlighting}[fontsize=\footnotesize]
\ExtensionTok{ParallelCluster}\NormalTok{ Template [default]: }\OperatorTok{<}\NormalTok{a name desired}\OperatorTok{>} 
\ExtensionTok{AWS}\NormalTok{ Access Key ID []: }\OperatorTok{<}\NormalTok{copy and paste the access key}\OperatorTok{>}
\ExtensionTok{AWS}\NormalTok{ Secret Access Key ID []: }\OperatorTok{<}\NormalTok{copy and paste the secret key}\OperatorTok{>}
\ExtensionTok{Acceptable}\NormalTok{ Values for AWS Region ID: }
    \ExtensionTok{eu-north-1}
    \ExtensionTok{ap-south-1}
    \ExtensionTok{eu-west-3}
    \ExtensionTok{eu-west-2}
    \ExtensionTok{eu-west-1}
    \ExtensionTok{ap-northeast-2}
    \ExtensionTok{ap-northeast-1}
    \ExtensionTok{sa-east-1}
    \ExtensionTok{ca-central-1}
    \ExtensionTok{ap-southeast-1}
    \ExtensionTok{ap-southeast-2}
    \ExtensionTok{eu-central-1}
    \ExtensionTok{us-east-1}
    \ExtensionTok{us-east-2}
    \ExtensionTok{us-west-1}
    \ExtensionTok{us-west-2}
\ExtensionTok{AWS}\NormalTok{ Region ID [ap-northeast-2]: }\OperatorTok{<}\NormalTok{the region to use}\OperatorTok{>}
\ExtensionTok{VPC}\NormalTok{ Name [}\OperatorTok{<}\NormalTok{default name}\OperatorTok{>}\NormalTok{]: }\OperatorTok{<}\NormalTok{a name desired}\OperatorTok{>}
\ExtensionTok{Acceptable}\NormalTok{ Values for Key Name: }
    \OperatorTok{<}\ExtensionTok{the}\NormalTok{ registered key names appear here}\OperatorTok{>}
\ExtensionTok{Key}\NormalTok{ Name []: }\OperatorTok{<}\NormalTok{enter the EC2 key pair name}\OperatorTok{>}
\ExtensionTok{Acceptable}\NormalTok{ Values for VPC ID: }
    \OperatorTok{<}\ExtensionTok{the}\NormalTok{ list of VPC appears here}\OperatorTok{>}
\ExtensionTok{VPC}\NormalTok{ ID []: }\OperatorTok{<}\NormalTok{enter one of the vpc above}\OperatorTok{>} 
\ExtensionTok{Acceptable}\NormalTok{ Values for Master Subnet ID: }
    \OperatorTok{<}\ExtensionTok{the}\NormalTok{ list of subnet ids appears here}\OperatorTok{>}
\ExtensionTok{Master}\NormalTok{ Subnet ID [subnet-}\OperatorTok{<}\NormalTok{default value}\OperatorTok{>}\NormalTok{]: }\OperatorTok{<}\NormalTok{enter one of the}
                          \NormalTok{master subnet id above}\OperatorTok{>}
\end{Highlighting}
\end{Shaded}

Now examine the files in the directory
\texttt{\textasciitilde{}/.parallelcluster} (a hidden directory under
the home directory). The file \texttt{pcluster-cli.log} shows the log
and the file \texttt{config} shows the configuration. One can modify the
file \texttt{config} to fine-tune the configuration per user's need.
The following is the \texttt{config} corresponding to our CfnCluster
experiments:

\begin{Verbatim}[fontsize=\footnotesize]
[global]
update_check = true
sanity_check = true
cluster_template = test

[aws]
aws_region_name = ap-northeast-2

[cluster test]
vpc_settings = testcfn
key_name = <key name>
initial_queue_size = 0
max_queue_size = 20
ebs_settings = expr_ebs
scheduler = sge
compute_instance_type = c5.18xlarge
master_instance_type = c5.18xlarge
cluster_type = spot
spot_price = 1.20
base_os = centos7
scaling_settings = custom
extra_json = {"cluster" : { "cfn_scheduler_slots" : "2"} }
master_root_volume_size = 20
compute_root_volume_size = 20

[ebs expr_ebs]
ebs_snapshot_id = < a snapshot id >
volume_size = 40

[vpc testcfn]
master_subnet_id = < a subnet id >
vpc_id = < a vpc id >

[aliases]
ssh = ssh {CFN_USER}@{MASTER_IP} {ARGS}

[scaling custom]
scaling_idletime = 20
\end{Verbatim}

In the \texttt{{[}global{]}} section, we set global configurations.
The \texttt{cluster\_template} names the cluster section to be used for the
cluster.\\
\texttt{update\_check} check for the updates to ParallelCluster, and
\texttt{sanity\_check} validates that resources defined in parameters.

In the \texttt{{[}aws{]}} section, the region is specified. AWS access key
and secret key may appear here unless specified in the base AWS CLI.

In the \texttt{{[}cluster{]}} section, we define the detailed specification
of the virtual cluster. The \texttt{vpc\_settings} names a setting for VPC,
detailed in the \texttt{{[}vpc{]}} section, and the \texttt{ebs\_settings} names
the setting for EBS, detailed in \texttt{{[}ebs{]}} section. The \texttt{key\_name} defines the key name to use.
The \texttt{initial\_queue\_size} defines the number of worker instances at
the launch of the cluster. We used zero for our experiments, as we often
needed to check if the configuration is done properly on master before
running actual jobs. The worker instances are launched upon submission
of a new job into the queue, and they are terminated when the workers
stay idle for a while (not exactly defined, but often around five to ten
minutes).

We set the \texttt{max\_queue\_size}, the maximum number of worker instances,
to 20. We used CentOS 7 as the \texttt{base\_os} for our instances.
The \texttt{master\_root\_volume\_size} and
the \texttt{compute\_root\_volume\_size} determine the size of root volume
of the master instance and each of the worker instance, respectively.
For the \texttt{scheduler}, we used the Sun Grid Engine (\texttt{sge}). 
For the \texttt{compute\_instance\_type}, we used \texttt{c5.18xlarge}, an
instance with 36 physical cores (72 virtual cores with hyperthreading).
It consists of two non-uniform memory access (NUMA) nodes with 18 physical cores each. 
In NUMA memory design, an access to local memory of a processor is faster than an access to non-local memory within a shared memory system. 
\texttt{master\_instance\_type} defines the instance type of the master.
Sometimes it is fine to be as small as \texttt{t2.micro}, a single-core
instance, but we needed an instance with good network performance when
many instances simultaneously accessed a large data file on shared
storage. The \texttt{cluster\_type} is either \texttt{ondemand} (default) or
\texttt{spot}. For \texttt{c5.18xlarge} in Seoul region
(\texttt{ap-northeast-2}), on-demand price was \$3.456 per instance-hour,
while the spot price was at \$1.0788 per instance-hour throughout the
duration of our experiments. Budget-constrained users may use spot instances for worker instances.
In case of this scenario,
the \texttt{spot\_prices} was set to \$1.20 per instance-hour, so if
the actual price went above this value, our worker instances would have
been terminated. Only the on-demand instance could be used as the master
instance, so smaller instance might be desirable for lower cost.
The setting
\texttt{extra\_json\ =\ \{"cluster"\ :\ \{\ "cfn\_scheduler\_slots"\ :\ "2"\}\ \}}
sets number of slots that an instance bears to two. Each computing job
is required to declare the number of ``slots'' to occupy. By default,
the number of slots per instance is the number of virtual cores the
instance has. This default setting is natural, but a problem arises if
we intend to utilize shared-memory parallelism in NUMA node-level, as
the number of slots occupied is tied to the number of instances 
launched. We assigned one slot per NUMA node that an instance has (i.e.,~2 slots
per instance) and utilized all 18 physical cores per NUMA node.

The \texttt{{[}ebs{]}} section defines the configuration for the EBS volume mounted
on the master node and shared via NFS to workers.
The \texttt{ebs\_snapshot\_id} defines the ID of the EBS snapshot to be
used. We had datasets and packages necessary for our jobs pre-installed
in an EBS volume and created a snapshot. The size of the volume
was 40 GB. By default, the volume is mounted to the path
\texttt{/shared}.

We refer the readers to
\url{https://docs.aws.amazon.com/parallelcluster/} for further details.

\subsection{Creating, accessing, and destroying the cluster}

We can create a virtual cluster named \texttt{example} by issuing the
following command on a local machine:

\begin{Shaded}
\begin{Highlighting}[fontsize=\footnotesize]
\ExtensionTok{pcluster}\NormalTok{ create example}
\end{Highlighting}
\end{Shaded}

To access the master instance through ssh, one needs the location of the
private key (\texttt{.pem}) file. The command to use is:

\begin{Shaded}
\begin{Highlighting}[fontsize=\footnotesize]
\ExtensionTok{pcluster}\NormalTok{ ssh example -i }\OperatorTok{<}\NormalTok{private key file}\OperatorTok{>}
\end{Highlighting}
\end{Shaded}

The default username for instances with CentOS is \texttt{centos}. The
default username depends on the Amazon Machine Image (AMI) being used to
create a virtual machine, which is determined by the \texttt{base\_os}
selected on the configuration.
The names of the existing clusters can be listed using the command
\texttt{pcluster\ list}, and we may completely remove a cluster
\texttt{example} using the command \texttt{pcluster\ delete\ example}.

\subsection{Installation of libraries}

Now we can access the master node through secure shell(SSH). We have a shared EBS
volume mounted at \texttt{/shared}, and we are to install necessary
software there. For our experiments, we installed anaconda, a portable installation of
Python, in the directory \texttt{/shared}. A script to set
up environment variables is also created and saved in \texttt{/shared}:

\begin{Shaded}
\begin{Highlighting}[fontsize=\footnotesize]
\CommentTok{# setup.sh}
\ExtensionTok{module}\NormalTok{ load mpi/openmpi-x86_64 }\CommentTok{# loads MPI to the environment}
\BuiltInTok{source}\NormalTok{ /shared/conda/etc/profile.d/conda.sh}
\BuiltInTok{export} \VariableTok{PATH=}\NormalTok{/shared/conda/bin:}\VariableTok{$PATH}
\BuiltInTok{export} \VariableTok{LD_LIBRARY_PATH=}\NormalTok{/shared/conda/lib:}\VariableTok{$LD_LIBRARY_PATH}
\end{Highlighting}
\end{Shaded}
We issued the command:
\begin{Shaded}
\begin{Highlighting}[fontsize=\footnotesize]
\BuiltInTok{source}\NormalTok{ setup.sh}
\end{Highlighting}
\end{Shaded}
to set up the environment variables.
We installed PyTorch from source\footnote{\url{https://github.com/pytorch/pytorch\#from-source}}, as it is required
to do so in order to incorporate MPI.

To download our code, one can issue the command:

\begin{Shaded}
\begin{Highlighting}[fontsize=\footnotesize]
\FunctionTok{git}\NormalTok{ clone https://github.com/kose-y/dist\_stat /shared/dist\_stat}
\end{Highlighting}
\end{Shaded}

\subsection{Running a job}

To provide instructions on how to define the environment to each
instance, we need a script defining each job. The following script
\texttt{mcpi-2.job} is for running the program for Monte Carlo
estimation of \(\pi\) in Section \ref{sec:lib} (Listing \ref{code:torch_mcpi}) using two instances (four
processes using 18 threads each).

\begin{Shaded}
\begin{Highlighting}[fontsize=\footnotesize]
\CommentTok{#!/bin/sh}
\CommentTok{#$ -cwd}
\CommentTok{#$ -N mcpi}
\CommentTok{#$ -pe mpi 4}
\CommentTok{#$ -j y}
\FunctionTok{date}
\BuiltInTok{source}\NormalTok{ /shared/conda/etc/profile.d/conda.sh}
\BuiltInTok{export} \VariableTok{PATH=}\NormalTok{/shared/conda/bin:}\VariableTok{$PATH}
\BuiltInTok{export} \VariableTok{LD_LIBRARY_PATH=}\NormalTok{/shared/conda/lib:}\VariableTok{$LD_LIBRARY_PATH}
\BuiltInTok{export} \VariableTok{MKL_NUM_THREADS=}\NormalTok{18}
\ExtensionTok{mpirun}\NormalTok{ -np 4 python /shared/dist\_stat/examples/mcpi-mpi-pytorch.py}
\end{Highlighting}
\end{Shaded}

The line \texttt{-pe\ mpi\ 4} tells the scheduler that we are using four
slots. Setting the value of the environment variable
\texttt{MKL\_NUM\_THREADS} to 18 means that MKL runs with 18 threads or
cores for that process. We launch four processes in the cluster, two per
instance, as defined by our ParallelCluster setup, in parallel using
MPI. We can submit this job to the Sun Grid Engine (the job scheduler)
using the command:

\begin{Shaded}
\begin{Highlighting}[fontsize=\footnotesize]
\ExtensionTok{qsub}\NormalTok{ mcpi-2.job}
\end{Highlighting}
\end{Shaded}

When we submit a job, a message similar to the following appears:

\begin{Shaded}
\begin{Highlighting}[fontsize=\footnotesize]
\ExtensionTok{Your}\NormalTok{ job 130 (}\StringTok{"mcpi"}\NormalTok{) }\ExtensionTok{has}\NormalTok{ been submitted}
\end{Highlighting}
\end{Shaded}

One may see the newly submitted job in the queue using
the command \texttt{qstat}.

\begin{Shaded}
\begin{Highlighting}[fontsize=\footnotesize]
\ExtensionTok{qstat}
\end{Highlighting}
\end{Shaded}

\begin{Shaded}
\begin{Highlighting}[fontsize=\scriptsize]
\ExtensionTok{job-ID}\NormalTok{  prior    name    user    state   submit/start at     queue    slots}
\ExtensionTok{------------------------------------------------------------------------------}
    \ExtensionTok{130}\NormalTok{ 0.55500  mcpi    centos  qw      02/28/2019 03:58:54              4 }
\end{Highlighting}
\end{Shaded}

If we want to delete any job waiting for the queue or running, use the
command \texttt{qdel}.

\begin{Shaded}
\begin{Highlighting}[fontsize=\footnotesize]
\ExtensionTok{qdel}\NormalTok{ 130}
\end{Highlighting}
\end{Shaded}

\begin{Shaded}
\begin{Highlighting}[fontsize=\footnotesize]
\ExtensionTok{centos}\NormalTok{ has deleted job 130}
\end{Highlighting}
\end{Shaded}

Once the job is completed, the output is saved as a text file named
such as \texttt{mcpi.o130}. For example:

\begin{Shaded}
\begin{Highlighting}[fontsize=\footnotesize]
\ExtensionTok{Thu}\NormalTok{ Feb 28 04:07:54 UTC 2019}
\ExtensionTok{3.148}
\end{Highlighting}
\end{Shaded}

The scripts for our numerical examples are in
\texttt{/shared/dist\_stat/jobs}.

\subsection{Miscellaneous}

To keep what is on the EBS volume on the cloud and access later, we need to 
create a snapshot for the volume. We can later create a volume based on
this snapshot\footnote{\url{https://docs.aws.amazon.com/AWSEC2/latest/UserGuide/ebs-creating-snapshot.html}}
and mount it on any instance\footnote{\url{https://docs.aws.amazon.com/AWSEC2/latest/UserGuide/ebs-using-volumes.html}}.
In ParallelCluster, this is done automatically when we give an ID of a
snapshot in the \texttt{config} file.
\section{Monte Carlo estimation of $\pi$ on multi-GPU using TensorFlow}\label{sec:mcpi_tf}

This section shows an implementation of Monte Carlo estimation example in Section \ref{sec:mcpi_torch} in multi-GPU using TensorFlow. 
Listing \ref{code:tf_mcpi} is the implementation that assumes a node with four GPUs. The code appears more or less the same as Listing \ref{code:torch_mcpi}, except that the list of devices is pre-specified. 
Line 2 indicates that a static computational graph is used (instead of the eager execution) for the function to run simultaneously on multiple GPUs.
It is slightly shorter than  Listing \ref{code:torch_mcpi}, partially because multi-GPU is supported natively in TensorFlow, and MPI is not used.

\begin{lstfloat}
\begin{lstlisting}[label=code:tf_mcpi,caption=Monte Carlo estimation of $\pi$ for TensorFlow on a workstation with multiple GPUs]
import tensorflow as tf

# Enforce graph computation. With eager execution, the code runs
# sequentially w.r.t. GPUs. e.g., computation for '/gpu:1' would not
# start until the computation for '/gpu:0' finishes.
@tf.function 
def mc_pi(n, devices):
    estim = []
    for d in devices:
        # use device d in this block
        with tf.device(d):
            x = tf.random.uniform((n,), dtype=tf.float64)
            y = tf.random.uniform((n,), dtype=tf.float64)
            # compute local estimate of pi 
            # and save it as an element of 'estim'.
            estim.append(tf.reduce_mean(tf.cast(x ** 2 +
                y ** 2 < 1, tf.float64)) * 4)
    return tf.add_n(estim)/len(devices)

if __name__ == '__main__':
    n = 10000
    devices = ['/gpu:0', '/gpu:1', '/gpu:2', '/gpu:3']
    r = mc_pi(n, devices)
    print(r.numpy())
\end{lstlisting}
\end{lstfloat}

\section{Further details on examples}\label{sec:numextra}
\subsection{Nonnegative matrix factorization}\label{sec:nmf}

The multiplicative algorithm discussed in \citet[Sect. 3.1]{zhou2010graphics} is due to \citet{lee1999learning,lee2001algorithms}: 
\begin{align*}
V^{n+1} &= V^n \odot [X (W^n)^T] \oslash [V^n W^n (W^n)^T] \\
W^{n+1} &= W^n \odot [(V^{n+1})^T X] \oslash [(V^{n+1})^T V^{n+1} W^n],
\end{align*}
where $\odot$ and $\oslash$ respectively denote elementwise multiplication and division.
This algorithm can be interpreted as an MM algorithm minimizing the surrogate function of $f(V, W)=\|X - VW\|_F^2$ based on Jensen's inequality:
\begin{align*}
g(V, W | V^n, W^n) = \sum_{i,j,k} \frac{v_{ik}^n w_{kj}^n}{\sum_{k'} v_{ik'}^n w_{k'j}^n} \left(x_{ij} - \frac{\sum_{k'} v_{ik'}^n w_{k'j}^n}{v_{ik}^n w_{kj}^n}v_{ik}w_{kj}\right)^2.
\end{align*}  
in an alternating fashion.

\begin{figure}
\centering
\includegraphics[width=0.3\textwidth]{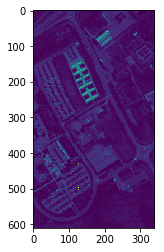}
\includegraphics[width=0.3\textwidth]{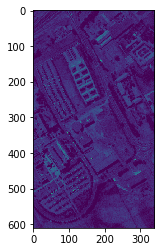}
\includegraphics[width=0.3\textwidth]{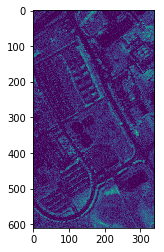}
\caption{Three selected bands from the NMF of the Pavia University hyperspectral image with $r=20$}\label{fig:nmf_pavia}
\end{figure}

Figure \ref{fig:nmf_pavia} shows an example of NMF on a publicly available hyperspectral image, acquired by the reflective optics system imaging spectrometer sensor in a flight campaign over Pavia University in Italy\footnote{Data available at \url{http://www.ehu.eus/ccwintco/index.php?title=Hyperspectral
Remote Sensing Scenes}.}. The image is essentially a $610\text{ (height)} \times 340\text{ (width)} \times 103\text{ (spactral bands)}$ cube. It is interpreted as a $207,400\text{ (pixels)} \times 103 \text{ (bands)}$ matrix and then analyzed using NMF. In the resulting $207,400 \times r$ matrix $V$, each column can be interpreted as a composite channel from the original 103 bands.
In the present experiments, the rank $r$ was set to 20.
Three of the twenty channels exhibiting distinct features were chosen by hand, and are shown in Figure \ref{fig:nmf_pavia}.

A problem with the multiplicative algorithm is the potential to generate subnormal numbers, which significantly slows down the algorithm. 
A subnormal (or denormal) number is a number smaller (in magnitude) than the smallest positive number that can be represented by the floating-point number system of the computer. Subnormal numbers are generated by the multiplicative algorithm if values smaller than 1 are multiplied repeatedly. 
Indeed, when Listing \ref{code:nmf} was run on a CPU with a small synthetic data of size $100 \times 100$, we observed a significant slowdown. 
The IEEE floating-point standard is to deal with subnormal numbers properly with a special hardware or software implementation \citep{ieee2008754}. 
In many CPUs, the treatment of subnormal numbers relies on software and hence is very slow. 
Forcing such values to zero is potentially dangerous 
because it is prone to the division-by-zero error. 
In contrast, Nvidia GPUs support subnormal numbers at a hardware level since the Fermi architecture, and simple arithmetic operations do not slow down by subnormal numbers \citep{whitehead2011precision}. 
Subnormal numbers can be completely avoided (especially in CPUs) by
using the APG method of the main text.

Distributing the multiplicative algorithm on a large scale machine is relatively straightforward \citep{liu2010distributed}. 
An implementation using \texttt{distmat} is given in Listing \ref{code:nmf}. The code for the APG algorithm of the main text is presented in Listing \ref{code:nmf2}.
%
In both Listings,
$X$ is assumed to be an $[m] \times p$ matrix. The resulting matrix $V$ is distributed as an $[m] \times r$ matrix, and $W$ is distributed as an $r \times [p]$ matrix. 
As discussed in Section \ref{sec:distmat}, distributed matrix multiplication algorithms are automatically selected from Table \ref{tab:mm} based on the arguments.

\begin{lstfloat}
\begin{lstlisting}[label=code:nmf,caption=A PyTorch code for multiplicative NMF update on a distributed system.]
import torch
import torch.distributed as dist
from dist_stat import distmat
from dist_stat.distmat import distgen_uniform

dist.init_process_group('mpi')
rank = dist.get_rank()
size = dist.get_world_size()
device = 'cuda:{}'.format(rank) # or simply 'cpu' for CPU computing
if device.startswith('cuda'): torch.cuda.set_device(rank)

# initialize X, W, V in a single device: a CPU or a GPU.
p = 10000
q = 10000
r = 20
max_iter=10000
TType = torch.cuda.FloatTensor if device.startswith('cuda') else torch.DoubleTensor
X = distgen_unifrom(p, q, TType=TType) # [p] x q
V = distgen_uniform(p, r, TType=TType) # [p] x r
W = distgen_uniform(q, r, TType=TType).t() # r x [q]

for i in range(max_iter):
    # Update V
    XWt =  distmat.mm(X, W.t()) # Scenario 1
    WWt =  distmat.mm(W, W.t()) # Scenario 5
    VWWt = distmat.mm(V, WWt)   # Scenario 4
    # V = V * XW^T / VWW^T elementwise. In-place operation.
    V = V.mul_(XWt).div_(VWWt) 
    
    # Update W
    VtX  = distmat.mm(V.t(), X, out_sizes=W.sizes) # Scenario 7
    VtV  = distmat.mm(V.t(), V)                    # Scenario 5
    VtVW = distmat.mm(VtV, W)                      # Scenario 11
    W = W.mul_(VtX).div_(VtVW)
    
    # compute objective value
    outer = distmat.mm(V, W)    # Scenario 2
    val = ((X - outer)** 2).all_reduce_sum()
    print(i, val.item())
\end{lstlisting}
\end{lstfloat}

\begin{lstfloat}
\begin{lstlisting}[label=code:nmf2,caption=A PyTorch code for alternating proximal gradient NMF update on a distributed system.]
import torch
import torch.distributed as dist
from dist_stat import distmat
from dist_stat.distmat import distgen_uniform

dist.init_process_group('mpi')
p = 10000
q = 10000
r = 20
eps = 1e-6
max_iter = 10000
TType = torch.cuda.FloatTensor if device.startswith('cuda') else torch.DoubleTensor
X = distgen_uniform(p, q, TType=TType) # [p] x q
V = distgen_uniform(p, r, TType=TType) # [p] x r
W = distgen_uniform(q, r, TType=TType).t() # r x [q]

for i in range(max_iter):
    XWt =  distmat.mm(X, W.t()) # Scenario 1
    WWt =  distmat.mm(W, W.t()) # Scenario 5
    VWWt = distmat.mm(V, WWt)   # Scenario 4
    sigma_k = 1.0/(2*((WWt**2).sum() + eps * r).sqrt())
    V = (V * (1.0 - sigma_k * eps) - (VWWt - XWt) * sigma_k).apply( \
        torch.clamp, min=0)

    VtX  = distmat.mm(V.t(), X, out_sizes=W.sizes) # Scenario 7
    VtV  = distmat.mm(V.t(), V)                    # Scenario 5
    VtVW = distmat.mm(VtV, W)                      # Scenario 11
    tau_k = 1.0/(2 * ((VtV ** 2).sum() + eps * r).sqrt())
    W = (W * (1.0 - tau_k * eps) - (VtVW - VtX) * tau_k).apply( \ 
        torch.clamp, min=0)

    outer = distmat.mm(V, W) # Scenario 2
    val = ((X - outer) ** 2).all_reduce_sum()
    print(i, val.item())
\end{lstlisting}
\end{lstfloat}

Table \ref{tab:nmf} shows the performance of the two NMF algorithms on a $[10,000] \times 10,000$ input matrix with various values of $r$. For APG, $\epsilon=0$ was used. 
While the APG algorithm required more operations per iteration than the multiplicative algorithm, it was faster on CPUs, because subnormal numbers were avoided. 
As they do not slow down with subnormal numbers, each iteration was faster in the multiplicative algorithm on GPUs. 
In Table \ref{tab:nmf_obj}, APG appears to converge slower early on (10,000 iterations), but eventually catches up the multiplicative algorithm (100,000 iterations) in terms of objective value. 
As more GPUs were used, the algorithms sped up. The only exception was with 8 GPUs with $r=60$, where inter-GPU communication overhead dominates the actual computation. 

Additional experiments were conducted to see how the value of $\epsilon$ affects the convergence. The results are shown in Table \ref{tab:nmf2}. Convergence was faster for higher values of $\epsilon$. 
The number of iterations to convergence in the multiplicative algorithm was larger than the APG with $\epsilon=10$ for higher-rank decompositions ($r$ = 40 and 60) due to heavier communication burden. 

In addition to ``small,'' 10,000 $\times$ 10,000 and ``large-size,'' 200,000 $\times$ 200,000 datasets used for the scalability experiment of the main text, we also considered the ``mid-size,'' 200,000 $\times$ 100,000 dataset. The results are shown in Table \ref{tab:nmf4}, complementing Table \ref{tab:nmf3} of the main text. As in the large-size dataset, the mid-size dataset did not fit in the memory of the eight GPUs.

\begin{table}[]
\caption{Runtime (in seconds) comparisons for NMF on the simulated $[10,000] \times 10,000$ data}\label{tab:nmf}
\centering
\begin{tabular}{lrrrrrr}
\hline
                   &     &\multicolumn{5}{c}{10,000 iterations} \\ \cline{3-7}
method             & $r$ & CPU  & 1 GPU & 2 GPUs & 4 GPUs & 8 GPUs \\ \hline
Multiplicative     & 20  & 655  & 160   & 93     & 62     & 50     \\
                   & 40  & 978  & 165   & 102    & 73     & 72     \\
                   & 60  & 1355 & 168   & 109    & 85     & 86     \\ \hline
APG                & 20  & 504  & 164   & 97     & 66     & 57     \\
($\epsilon = 0$)   & 40  & 783  & 168   & 106    & 78     & 77     \\
                   & 60  & 1062 & 174   & 113    & 90     & 92     \\ \hline
\end{tabular}
\end{table}

\begin{table}[]
\caption{Comparison of objective function values for simulated $[10,000] \times 10,000$ data after 10,000 iterations and 100,000 iterations}\label{tab:nmf_obj}
\centering
\begin{tabular}{lrrr}
\hline
method & $r$ & 10,000 iterations & 100,000 iterations \\ \hline
Multiplicative     & 20  &  8.270667E+06 & 8.270009E+06 \\
                   & 40  &  8.210266E+06 & 8.208682E+06 \\
                   & 60  &  8.155084E+06 & 8.152358E+06 \\ \hline
APG                & 20  &  8.271248E+06 & 8.270005E+06 \\
($\epsilon = 0$)   & 40  &  8.210835E+06 & 8.208452E+06 \\
                   & 60  &  8.155841E+06 & 8.151794E+06 \\ \hline
\end{tabular}
\end{table}

\begin{sidewaystable}[]
\caption{Convergence time comparisons for different values of $\epsilon$ in APG and the multiplicative method} \label{tab:nmf2}
\centering
\begin{tabular}{lrrrrrrrrrrr}
\hline
& \multicolumn{3}{c}{$r=20$, 8 GPUs} & & \multicolumn{3}{c}{$r=40$, 8 GPUs} & &\multicolumn{3}{c}{$r=60$, 4 GPUs} \\
\cline{2-4}\cline{6-8}\cline{10-12}
Method             & iterations & time (s) & function     & & iterations & time (s) & function     & & iterations & time (s) & function     \\ \hline
Multiplicative     & 21200      & 110      & 8.270530E+06 & & 36600      & 269      & 8.209031E+06 & & 50000      & 446      & 8.152769E+06 \\
APG $\epsilon=0$   & 31500      & 198      & 8.270202E+06 & & 37400      & 310      & 8.208875E+06 & & 55500      & 536      & 8.152228E+06 \\
APG $\epsilon=0.1$ & 30700      & 191      & 8.274285E+06 & & 36700      & 302      & 8.210324E+06 & & 55500      & 537      & 8.153890E+06 \\
APG $\epsilon=1$   & 30500      & 190      & 8.282346E+06 & & 37300      & 307      & 8.223108E+06 & & 47800      & 460      & 8.168503E+06 \\
APG $\epsilon=10$  & 28000      & 178      & 8.389818E+06 & & 31000      & 257      & 8.347859E+06 & & 46400      & 448      & 8.308998E+06 \\ \hline
\end{tabular}
\caption{Runtime of APG algorithm for NMF on simulated data for 1000 iterations on AWS EC2 \texttt{c5.18xlarge} instances.}
\centering
\begin{tabular}{@{\extracolsep{10pt}}rrrr@{}}
\hline 
configuration   & \multicolumn{3}{c}{200,000 $\times$ 100,000}        \\
instances   & $r=20$                 & $r=40$ & $r=60$  \\ \hline
1 &$\times$ &$\times$ &$\times$ \\
2 &  1487                   & 1589   & 2074         \\
4  &  863                    & 998    & 1233       \\
5  &  661                    & 896    & 1082       \\
8  &  448                    & 541    & 688        \\
10 &  422                    & 540    & 682        \\
20 & 363                    & 489    & 592       \\ \hline 
\end{tabular}\label{tab:nmf4}
\end{sidewaystable}

\subsection{Positron emission tomography}\label{sec:pet}
\subsubsection*{MM algorithm}
Recall that the log likelihood of the PET reconstruction problem is
\begin{align*}
L(\lambda) &= \sum_{i=1}^d \Big[ y_i \log \big(\sum_{j=1}^p e_{ij}\lambda_j \big) - \sum_{j=1}^p e_{ij} \lambda_j\Big].
\end{align*}
%
The MM algorithm considered in \citet[Sect. 3.2]{zhou2010graphics} for ridge-penalized maximum likelihood estimation of the intensity map $\lambda$ maximizes $L(\lambda) - (\mu/2)\|D\lambda\|_2^2$
based on separation of the penalty function by the minorization 
$$
(\lambda_j - \lambda_k)^2 \ge - \frac{1}{2} (2 \lambda_j - \lambda_{j}^n - \lambda_{k}^n)^2 - \frac{1}{2} (2 \lambda_k - \lambda_{j}^n - \lambda_{k}^n)^2
$$
(recall that each row of $D$ has one +1 and one $-1$), 
yielding iteration
\begin{align*}
z_{ij}^{n+1} &= e_{ij} y_i \lambda_{j}^n / \big(\sum_k e_{ik} \lambda_{k}^n\big)  \\
b_{j}^{n+1}  &= \mu \big(n_j \lambda_{j}^n + \sum_{k} g_{jk} \lambda_{k}^n\big) -1  \\
\lambda_{j}^{n+1} &= \Big(-b_{j}^{n+1} - \big[(b_{j}^{n+1})^2 - 4 a_j  \sum_{i=1}^d z_{ij}^{n+1}\big]^{1/2}\Big) / (2 a_j), 
\end{align*}
where
$n_j$ is the degree of the pixel $j$ on the pixel grid, and $a_j = -2 \mu n_j$.
By using matrix notation, the \texttt{distmat} implementation of this algorithm can be succinctly written as in Listing \ref{code:pet}. 

\begin{lstfloat}
\begin{lstlisting}[label=code:pet,caption=PyTorch code for PET with a squared difference penalty.]
import torch
import torch.distributed as dist
from dist_stat import distmat
from dist_stat.distmat import distgen_ones

dist.init_process_group('mpi')
rank = dist.get_rank()
size = dist.get_world_size()
device = 'cuda:{}'.format(rank) # or simply 'cpu' for CPU computing
if device.startswith('cuda'): torch.cuda.set_device(rank)

# (Data reading part omitted.)
# G: adjacency matrix, sparse [p] x p.
# E: detection probability matrix, d x [p].
# D: difference matrix, #edges x [p].
# y: observed count data, d x 1 broadcast.

TType = torch.cuda.FloatTensor if device.startswith('cuda') else torch.DoubleTensor
eps = 1e-20 # a small number for numerical stability
lambd = distmat.distgen_ones(G.shape[0], 1).type(TType) # poisson intensity, [p] x 1.
mu = 1e-6 # roughness penalty parameter
# compute |N_j|, row-sums of G.
N = distmat.mm(G, torch.ones(G.shape[1], 1).type(TType)) # [p] x 1.
a = -2 * mu * N # [p] x 1.
el = distmat.mm(E, lambd) # Scenario 5. output d x 1.
gl = distmat.mm(G, lambd) # Scenario 1. output [p] x 1.

for i in range(max_iter):
    z  = E * y * lambd.t() / (el + eps) # d x [p].
    b = mu * (N * lambd + gl) -1 # [p] x 1.
    c = z.sum(dim=0).t() # [p] x 1.
    # update lambda
    if mu != 0:
        lambd = (-b - (b ** 2 - 4 * a * c).sqrt()) / (2 * a + eps) # [p] x 1.
    else:
        lambd = -c / (b + eps) # [p] x 1.
    el = distmat.mm(E, lambd) # Scenario 5. output d x 1.
    gl = distmat.mm(G, lambd) # Scenario 1. output [p] x 1.

    likelihood = (y * torch.log(el + eps) - el).sum()
    dl = distmat.mm(D, lambd) # Scenario 5. output #edges x 1.
    penalty = - mu / 2.0 * torch.sum(dl ** 2)
    print(i, likelihood + penalty)
\end{lstlisting}
\end{lstfloat}

\begin{figure}[t!]
\centering
\begin{subfigure}[t]{0.22\textwidth}
\includegraphics[width=\textwidth]{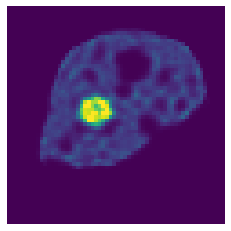}
\caption{$\mu=0$}
\end{subfigure}
\begin{subfigure}[t]{0.22\textwidth}
\includegraphics[width=\textwidth]{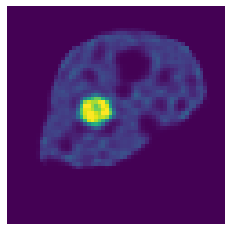}
\caption{$\mu=10^{-6}$}
\end{subfigure}
\begin{subfigure}[t]{0.22\textwidth}
\includegraphics[width=\textwidth]{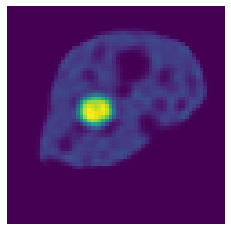}
\caption{$\mu=10^{-5}$}
\end{subfigure}
\begin{subfigure}[t]{0.22\textwidth}
\includegraphics[width=\textwidth]{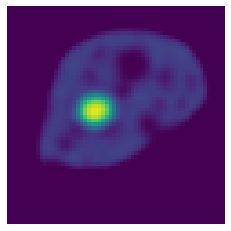}
\caption{$\mu=10^{-4}$}
\end{subfigure}
\caption{Reconstructed images of the 
XCAT phantom with a ridge penalty.}
\label{fig:pet_l2_xcat}
\end{figure}

Figure \ref{fig:pet_l2_xcat} shows the results of applying this iteration to the XCAT phantom of the main text. 
Images get smooth as the value of $\mu$ increases, but the edges get blurry.
Compare the edge contrast with Figure \ref{fig:pet_l1_xcat} (panels a--d) of the main text, in which anisotropic TV penalty was used and the estimation was conducted by using the PDHG algorithm.
The \texttt{distmat} implementation of the latter algorithm is given in Listing \ref{code:pet_l1}.
Table \ref{tab:pet2} shows the convergence behavior of PDHG with different values of penalty parameters. Observe that the algorithm converges faster for large values of $\rho$ with which the solution gets sparse. 

\begin{lstfloat}
\begin{lstlisting}[basicstyle=\scriptsize\ttfamily,label=code:pet_l1,caption=PyTorch code for PET with absolute value penalty.]
import torch
import torch.distributed as dist
from dist_stat import distmat
from dist_stat.distmat import distgen_ones

dist.init_process_group('mpi')
rank = dist.get_rank()
size = dist.get_world_size()
device = 'cuda:{}'.format(rank) # or simply 'cpu' for CPU computing
if device.startswith('cuda'): torch.cuda.set_device(rank)

# (Data reading part omitted.)
# E: detection probability matrix, d x [p].
# D: difference matrix, #edges x [p].
# y: observed count data, d x 1 broadcast.

TType = torch.cuda.FloatTensor if device.startswith('cuda') else torch.DoubleTensor
Et, Dt = E.t(), D.t() # transpose

lambd = distmat.distgen_ones(E.shape[1], 1).type(TType) # poisson intensity, [p] x 1.
lambd_prev = distmat.disten_ones(E.shape[1], 1).type(TType)
lambd_tilde = distmat.disten_ones(E.shape[1], 1).type(TType)

eps = 1e-20 # a small number for numerical stability
rho = 1e-4 # penalty parameter
tau = 1/3 # primal step size
sig = 1/3 # dual step size

w = torch.zeros(D.shape[0], 1).type(TType) # #edges x 1.
z = - torch.ones(E.shape[0], 1).type(TType) # d x 1.
Et1 = - distmat.mm(Et, z) # Scenario 4, [p] x 1.

for i in range(max_iter):
    # dual update
    el = distmat.mm(E, lambd_tilde) # Scenario 5, d x 1.
    z = z + sig * el # d x 1
    tmp = (z ** 2 + 4 * sig * y).sqrt() # d x 1
    z = 0.5 * (z - tmp) # d x 1
    dl = distmat.mm(D, lambd_tilde) # Scenario 5, #edges x 1.
    w = w + sig * dl # #edges x 1
    w = torch.clamp(w, max=rho, min=-rho) # #edges x 1
    
    # primal update
    Etz = distmat.mm(Et, z) # Scenario 4, [p] x 1.
    Dtw = distmat.mm(Dt, w) # Scenario 4, [p] x 1.
    lambd_prev = lambd
    lambd = lambd - tau * (Etz + Dtw + Et1) # [p] x 1.
    lambd = lambd.apply(torch.clamp, min=0.0) # [p] x 1.
    lambd_tilde = 2 * lambd - lambd_prev

    # objective
    el = distmat.mm(E, lambd) # Scenario 5, d x 1.
    dl = distmat.mm(D, lambd) # Scenario 5, #edges x 1.
    likelihood = (y * torch.log(el + eps) - el).sum()
    penalty = - rho * torch.sum(dl.abs())
    print(i, likelihood + penalty)
\end{lstlisting}
\end{lstfloat}

\begin{figure}[h!]
\centering
\begin{subfigure}[b]{0.3\textwidth}
\includegraphics[width=\textwidth]{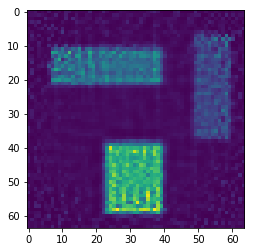}
\caption{$\mu=0$}
\end{subfigure}
\begin{subfigure}[b]{0.3\textwidth}
\includegraphics[width=\textwidth]{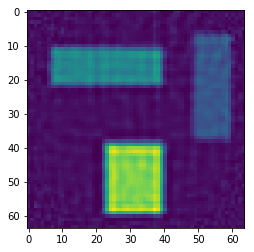}
\caption{$\mu=10^{-7}$}
\end{subfigure}\\
\begin{subfigure}[b]{0.3\textwidth}
\includegraphics[width=\textwidth]{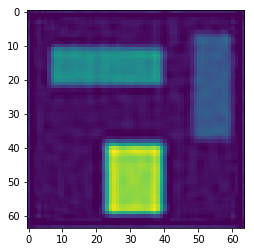}
\caption{$\mu=10^{-6}$}
\end{subfigure}
\begin{subfigure}[b]{0.3\textwidth}
\includegraphics[width=\textwidth]{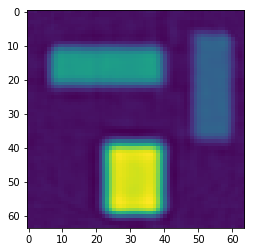}
\caption{$\mu=10^{-5}$}
\end{subfigure}
\caption{Reconstructed images of the 
RVF phantom with a ridge penalty.}
\label{fig:pet_l2}
\centering
\begin{subfigure}[b]{0.3\textwidth}
\includegraphics[width=\textwidth]{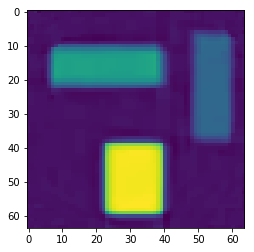}
\caption{$\rho=2^{-10}$}
\end{subfigure}
\begin{subfigure}[b]{0.3\textwidth}
\includegraphics[width=\textwidth]{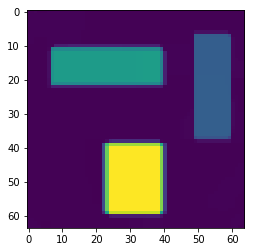}
\caption{$\rho=2^{-8}$}
\end{subfigure}\\
\begin{subfigure}[b]{0.3\textwidth}
\includegraphics[width=\textwidth]{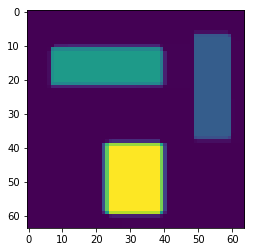}
\caption{$\rho=2^{-6}$}
\end{subfigure}
\begin{subfigure}[b]{0.3\textwidth}
\includegraphics[width=\textwidth]{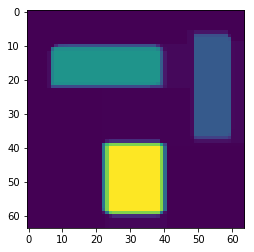}
\caption{$\rho=2^{-4}$}
\end{subfigure}
\caption{Reconstructed images of the 
RVF phantom with a TV penalty.}
\label{fig:pet_l1}
\end{figure}

To complete the comparison with \citet{zhou2010graphics}, we present here 
Figures \ref{fig:pet_l2} and \ref{fig:pet_l1} that show similar results with the $p= 64 \times 64$ Roland-Varadhan-Frangakis phantom \citep{roland2007squared} with $d = 2016$, used in the cited paper, for various values of regularization parameters $\mu$ and $\rho$. 

\begin{table}[]
\caption{Convergence time comparisons for TV-penalized PET with different values of $\rho$. Problem dimension is $p=10,000$ and $d = 16,110$. Eight GPUs were used.}\label{tab:pet2}
\centering
\begin{tabular}{rrrr}
\hline
$\rho$ & iterations & time (s) & function     \\ \hline
0      & 6400       & 20.6     & -2.417200E+05 \\
0.01   & 4900       & 15.8     & -2.412787E+05 \\
0.1    & 5000       & 16.1     & -2.390336E+05 \\
1      & 2800       & 9.5      & -2.212579E+05 \\ \hline
\end{tabular}
\end{table}

\subsubsection*{Stochastic PDHG}
The stochastic PDHG algorithm employed to produce the bottom row of Figure \ref{fig:pet_l1_xcat} is based on the dual version of iteration \eqref{eqn:cp} \citep{Condat:JournalOfOptimizationTheoryAndApplications:2012,ko2019easily}:
\begin{align*}
x^{n+1} &= \mathbf{prox}_{\tau g} (x^n - \tau K^T y^{n+1})  \\
y^{n+1} &= \mathbf{prox}_{\sigma f^*}(y^n + \sigma K \bar{x}^n)  \\
\bar{y}^{n+1} &= 2y^{n+1}- y^n.
\end{align*}
If we take $f(z, w)=\sum_{i=1}^d f_i(z_i) + \rho\|w\|_1$ with $f_i(z) = - y_i \log z$,
$g(\lambda) = \mathbf{1}^T E \lambda + \delta_+ (\lambda)$, 
and $K = [E_1^T, \dotsc, E_d^T, D^T]^T$ where $E_i$ is the $i$th row of the mixing matrix $E$, then the above iteration reduces to
\begin{align*}
\lambda^{n+1} &= P_+ (\lambda^n - \tau (E^T \bar{z}^n + D^T \bar{w}^n + E^T \mathbf{1})) \\
z^{n+1}_i &= 
\big[(z_i^n + \sigma E_i{\lambda}^{n+1}) - [(z_i^n + \sigma E_i\lambda^{n+1})^2 + 4 \sigma y_i]^{1/2}\big]/2,
\quad i=1, \dotsc, d
\\
w^{n+1} &= P_{[-\rho, \rho]}(w^n + \sigma D \lambda^{n+1}) \\
\bar{z}^{n+1} &= 2z^{n+1} - z^{n} \\
\bar{w}^{n+1} &= 2 w^{n+1} - w^{n}.
\end{align*}
\citet{chambolle2018} propose to replace the $\tilde{z}$-update by sampling and then debias the $z$-update accordingly. If uniform sampling is used, then the second and fourth lines are replaced by:
\begin{align*}
z_i^{n+1} &= \begin{cases}
\big[(z_i^n + \sigma E_i{\lambda}^{n+1}) - [(z_i^n + \sigma E_i\lambda^{n+1})^2 + 4 \sigma y_i]^{1/2}\big]/2, & 
\text{with probability $\pi$} \\
\tilde{z}^{n}_i, & \text{otherwise} 
\end{cases} \\
\bar{z}^{n+1} &= z^{n+1} + \pi^{-1}(z^{n+1} - z^{n}).
\end{align*}
In the experiments for Figure \ref{fig:pet_l1_xcat} (panels e--h), $\pi$=0.2 was used. That is, only 20\% of the coordinates were updated 
every iteration.

\subsection{Multidimensional scaling}\label{sec:mds}
The parallel MM iteration of the main text 
\begin{align*}
\theta_{ik}^{n+1} &= \big(\sum_{j \ne i} \big[y_{ij}\textstyle\frac{\theta_{ik}^n - \theta_{jk}^n}{\|\theta_{i}^n - \theta_{j}^n\|_2} + (\theta_{ik}^n + \theta_{jk}^n)\big]\big) \big/ \big(2 \sum_{j \ne i} w_{ij} \big)
\end{align*}
is derived from minimizing the surrogate function
\begin{align*}
g(\theta | \theta^n) &= 2 \sum_{i=1}^q \sum_{j \ne i} \left[w_{ij}\left\|\theta_{i} - \frac{1}{2} (\theta_{i}^n+\theta_{j}^n)\right\|_2^2 - \frac{w_{ij}y_{ij} (\theta_i)^T (\theta_{i}^n - \theta_{j}^n)}{\|\theta_{i}^n - \theta_{j}^n\|_2}\right].
\end{align*}
that majorizes the stress function $f(\theta)=\sum_{i =1}^q \sum_{j \ne i} w_{ij}(y_{ij} - \|\theta_i - \theta_j \|_2)^2$

In PyTorch syntax, this can be computed in parallel using the code in Listing \ref{code:mds}, when all the data points are equally weighted.
\begin{lstfloat}[ht]
\begin{lstlisting}[label=code:mds,caption=PyTorch code for MDS.]
import torch
import torch.distributed as dist
from dist_stat import distmat
from dist_stat.distmat import distgen_normal, distgen_uniform
from dist_stat.application.euclidean_distance import euclidean_distance_DistMat
from math import inf

dist.init_process_group('mpi')
rank = dist.get_rank()
size = dist.get_world_size()
device = 'cuda:{}'.format(rank) # or simply 'cpu' for CPU computing
if device.startswith('cuda'): torch.cuda.set_device(rank)

m = 10000
d = 10000
q = 20
max_iter = 10000
TType = torch.cuda.FloatTensor if device.startswith('cuda') else torch.DoubleTensor
X = distgen_normal(m, d, TType=TType) # [m] x d
y = euclidean_distance_DistMat(X).type(TType) # [m] x m
w_sums = float(m - 1.0)
theta = distgen_uniform(m, q, lo=-1, hi=1, TType=TType) # [m] x q

for i in range(max_iter):
    d = distmat.mm(theta, theta.t()) # Scenario 2.
    TtT_diag_nd = d.diag(distribute=False).view(1, -1) # local
    TtT_diag    = d.diag(distribute=True) # distributed col vector
    d = d.mul_(-2.0)
    d.add_(TtT_diag_nd)
    d.add_(TtT_diag)

    d.fill_diag_(inf)
    Z = distmat.div(y, d)
    Z_sums = Z.sum(dim=1) # length-q broadcast vector
    WmZ = distmat.sub(1.0, Z)
    WmZ.fill_diag_(0.0)
    TWmZ = distmat.mm(theta.t(), WmZ.t()) # Scenario 8
    theta = (theta * (w_sums + Z_sums) + TWmZ.t()) / (w_sums * 2.0)

    distances = euclidean_distance_DistMat(theta)
    obj = ((y - distances)**2).sum() / 2.0
    print(i, obj.item())
\end{lstlisting}
\end{lstfloat}
In the numerical experiments of Sect. \ref{sec:zhou} of the main text, the pairwise Euclidean distances between data points were computed distributedly  \citep{li2010chunking}: 
in each stage, data on one of the processors are broadcast and each processor computes pairwise distances between the data residing on its memory and the broadcast data. This is repeated until all the processors broadcast its data. Note that the dimension of the datapoints does not matter after computing the pairwise distances.

\section{Details of SNPs selected in $\ell_1$-regularized Cox regression}\label{sec:snps}
\begin{figure}[h!]
\centering
\includegraphics[width=0.8\textwidth]{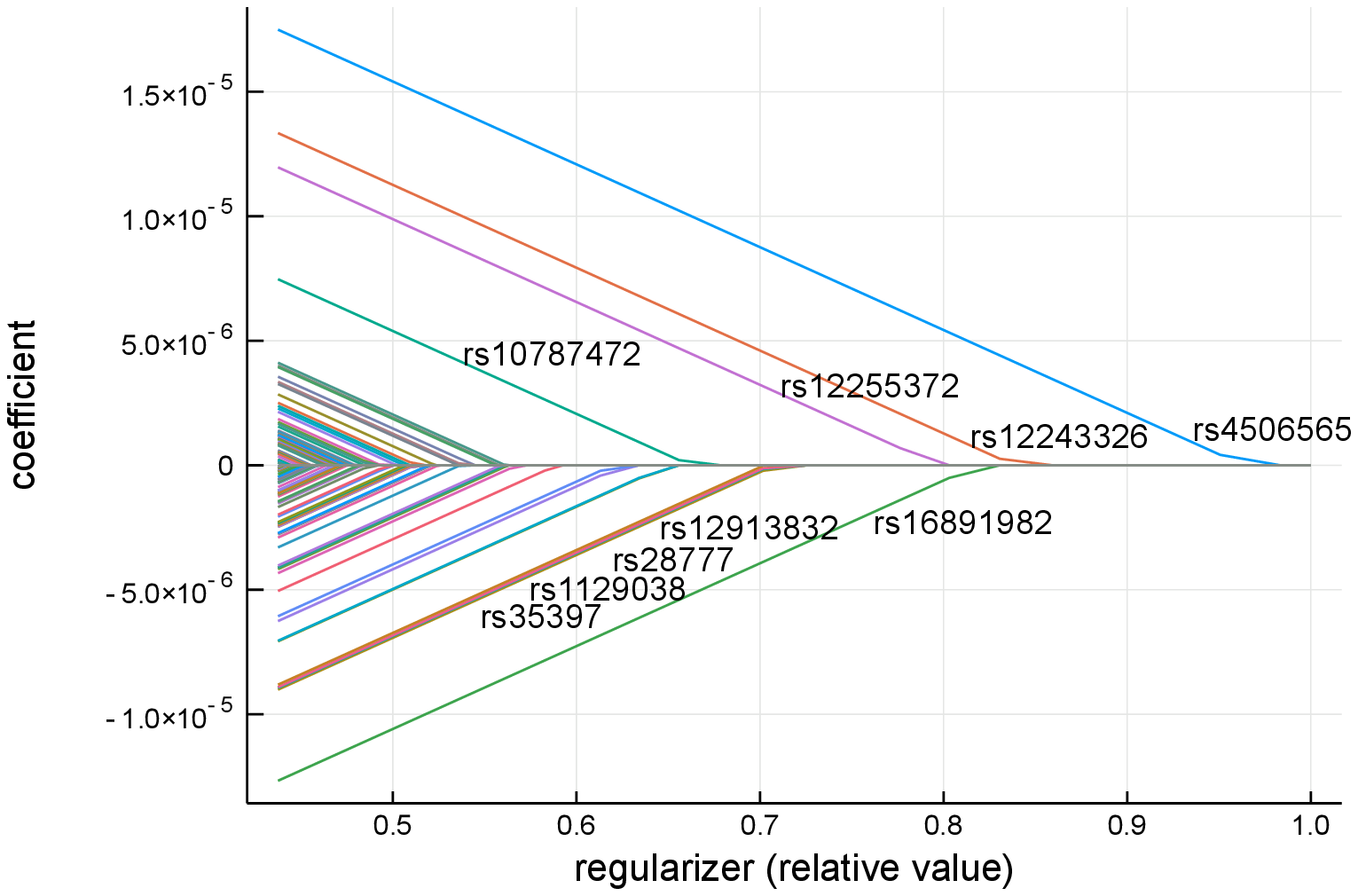}
\caption{Solution path for $\ell_1$-regularized Cox regression on the UK Biobank dataset. Signs are with respect to the reference allele: positive value favors alternative allele as the risk allele.}\label{fig:path}
\end{figure}

Figure \ref{fig:path} shows the solution path for SNPs within the range we used for the experiment in Section \ref{sec:biobank}. Tables \ref{tab:snps_pt1} and \ref{tab:snps_pt2} list the 111 selected SNPs. 

\begin{table}[h!]
\centering
\resizebox{\textwidth}{!}{
\begin{threeparttable}
\caption{SNPs selected by $\ell_1$-penalized Cox regression: \#1-\#56}\label{tab:snps_pt1}
\begin{tabular}{rlrrllrlll}
\hline
Rank & SNP ID     & Chr\tnote{A} & Location  & A1\tnote{B}  & A2\tnote{C} & MAF\tnote{D}   & Mapped genes                       & Sign\tnote{E} & Known\tnote{F} \\ \hline
1  & rs4506565  & 10 & 114756041 & A   & \textbf{T} & 0.238 & TCF7L2                 & $+$ & Yes \\
2  & rs12243326 & 10 & 114788815 & \textbf{C}   & T & 0.249 & TCF7L2                 & $+$ & Yes \\
3  & rs16891982 & 5  & 33951693  & G   & \textbf{C} & 0.215 & SLC45A2                & $-$ &     \\
4  & rs12255372 & 10 & 114808902 & \textbf{T}   & G & 0.215 & TCF7L2                 & $+$ & Yes \\
5  & rs12913832 & 15 & 28365618  & G   & \textbf{A} & 0.198 & HERC2                  & $-$ &     \\
6  & rs28777    & 5  & 33958959  & \textbf{C}   & A & 0.223 & SLC45A2                & $-$ &     \\
7  & rs1129038  & 15 & 28356859  & \textbf{C}   & T & 0.343 & HERC2                  & $-$ &     \\
8  & rs35397    & 5  & 33951116  & T   & \textbf{G} & 0.304 & SLC45A2                & $-$ &     \\
9  & rs10787472 & 10 & 114781297 & \textbf{C}   & A & 0.430 & TCF7L2                 & $+$ & Yes \\
10 & rs2470890  & 15 & 75047426  & T   & \textbf{C} & 0.429 & CYP1A2                 & $-$ &     \\
11 & rs2472304  & 15 & 75044238  & A   & \textbf{G} & 0.460 & CYP1A2                 & $-$ &     \\
12 & rs1378942  & 15 & 75077367  & A   & \textbf{C} & 0.401 & CSK, MIR4513           & $-$ &     \\
13 & rs34862454 & 15 & 75101530  & T   & \textbf{C} & 0.416 & LMAN1L                 & $-$ &     \\
14 & rs849335   & 7  & 28223990  & C   & \textbf{T} & 0.406 & JAZF1, JAZF1-AS1       & $-$ & Yes \\
15 & rs864745   & 7  & 28180556  & C   & \textbf{T} & 0.316 & JAZF1                  & $-$ & Yes \\
16 & rs12785878 & 11 & 71167449  & T   & \textbf{G} & 0.251 & NADSYN1, DHCR7         & $-$ &     \\
17 & rs4944958  & 11 & 71168073  & G   & \textbf{A} & 0.237 & NADSYN1, DHCR7         & $-$ &     \\
18 & rs8042680  & 15 & 91521337  & \textbf{A}   & C & 0.277 & PRC1, PRC1-AS1, Y\_RNA & $+$ &     \\
19 & rs35414    & 5  & 33969628  & \textbf{T}   & C & 0.188 & SLC45A2                & $-$ &     \\
20 & rs1635852  & 7  & 28189411  & \textbf{T}   & C & 0.423 & JAZF1                  & $-$ &     \\
21 & rs10962525 & 9  & 16659863  & \textbf{T}   & C & 0.321 & BNC2                   & $+$ &     \\
22 & rs1446585  & 2  & 136407479 & \textbf{G}   & A & 0.322 & R3HDM1                 & $+$ &     \\
23 & rs7570971  & 2  & 135837906 & \textbf{A}   & C & 0.327 & RAB3GAP1               & $+$ &     \\
24 & rs36074798 & 15 & 91518800  & \textbf{ACT} & A & 0.328 & PRC1, PRC1-AS1, Y\_RNA & $+$ & Yes \\
25 & rs10962612 & 9  & 16804167  & G   & \textbf{T} & 0.088 & BNC2                   & $-$ &     \\
26 & rs10962612 & 2  & 135911422 & T   & \textbf{C} & 0.097 & RAB3GAP1, ZRANB3       & $+$ &     \\
27 & rs941444   & 17 & 17693891  & \textbf{C}   & G & 0.073 & RAI1                   & $-$ & Yes \\
28 & rs6769511  & 3  & 185530290 & T   & \textbf{C} & 0.045 & IGF2BP2                & $+$ & Yes \\
29 & rs916977   & 15 & 28513364  & \textbf{T}   & C & 0.044 & HERC2                  & $-$ &     \\
30 & rs35390    & 5  & 33955326  & \textbf{C}   & A & 0.062 & SLC45A2                & $-$ &     \\
31 & rs35391    & 5  & 33955673  & \textbf{T}   & C & 0.374 & SLC45A2                & $-$ &     \\
32 & rs1470579  & 3  & 185529080 & A   & \textbf{C} & 0.436 & IGF2BP2                & $+$ & Yes \\
33 & rs2862954  & 10 & 101912064 & \textbf{T}   & C & 0.488 & ERLIN1                 & $-$ &     \\
34 & rs2297174  & 9  & 16706557  & A   & \textbf{G} & 0.346 & BNC2                   & $-$ &     \\
35 & rs1667394  & 15 & 28530182  & T   & \textbf{C} & 0.274 & HERC2                  & $-$ &     \\
36 & rs12440952 & 15 & 74615292  & \textbf{G}   & A & 0.279 & CCDC33                 & $+$ &     \\
37 & rs56343038 & 9  & 16776792  & G   & \textbf{T} & 0.318 & BNC2, LSM1P1           & $-$ &     \\
38 & rs9522149  & 13 & 111827167 & \textbf{T}   & C & 0.395 & ARHGEF7                & $-$ &     \\
39 & rs343092   & 12 & 66250940  & \textbf{T}   & G & 0.463 & HMGA2, HMGA2-AS1       & $-$ & Yes \\
40 & rs10733316 & 9  & 16696626  & \textbf{T}   & C & 0.436 & BNC2                   & $-$ &     \\
41 & rs823485   & 1  & 234671267 & T   & \textbf{C} & 0.488 & LINC01354              & $+$ &     \\
42 & rs12910825 & 15 & 91511260  & A   & \textbf{G} & 0.384 & PRC1, PRC1-AS1, RCCD1  & $+$ & Yes \\
43 & rs2959005  & 15 & 74618128  & T   & \textbf{C} & 0.222 & CCDC33                 & $-$ &     \\
44 & rs10756801 & 9  & 16740110  & T   & \textbf{G} & 0.494 & BNC2                   & $-$ &     \\
45 & rs12072073 & 1  & 3130016   & \textbf{C}   & T & 0.497 & PRDM16                 & $+$ &     \\
46 & rs7039444  & 9  & 20253425  & T   & \textbf{C} & 0.360 & (intergenic variant)   & $+$ &     \\
47 & rs7899137  & 10 & 76668462  & \textbf{A}   & C & 0.289 & KAT6B                  & $-$ &     \\
48 & rs11078405 & 17 & 17824978  & \textbf{T}   & G & 0.291 & TOM1L2                 & $+$ &     \\
49 & rs830532   & 5  & 142289541 & C   & \textbf{T} & 0.333 & ARHGAP26               & $+$ &     \\
50 & rs833283   & 3  & 181590598 & \textbf{G}   & C & 0.352 & (intergenic variant)   & $-$ &     \\
51 & rs10274928 & 7  & 28142088  & \textbf{A}   & G & 0.365 & JAZF1                  & $-$ & Yes \\
52 & rs13301628 & 9  & 16665850  & A   & \textbf{C} & 0.412 & BNC2                   & $-$ &     \\
53 & rs885107   & 16 & 30672719  & \textbf{C}   & T & 0.353 & PRR14, FBRS            & $+$ &     \\
54 & rs8180897  & 8  & 121699907 & A   & \textbf{G} & 0.445 & SNTB1                  & $+$ &     \\
55 & rs23282    & 5  & 142270301 & G   & \textbf{A} & 0.225 & ARHGAP26               & $+$ &     \\
56 & rs6428460  & 1  & 198377460 & C   & \textbf{T} & 0.229 & (intergenic variant)   & $+$ &           \\ \hline
\end{tabular}
\begin{tablenotes}
\item[A] Chromosome,
\item[B] Minor allele,
\item[C] Major allele,
\item[D] Minor allele frequency,
\item[E] Sign of the regression coefficient,
\item[F] Mapped gene included in \citet{mahajan2018fine}.
The boldface indicates the risk allele determined by the reference allele and the sign of the regression coefficient. 
\end{tablenotes}
\end{threeparttable}
}
\end{table}

\begin{table}[h!]
\centering
\resizebox{\textwidth}{!}{
\begin{threeparttable}
\caption{SNPs selected by $\ell_1$-penalized Cox regression: \#57-\#111}\label{tab:snps_pt2}
\begin{tabular}{rlrrllrlll}
\hline
Rank & SNP ID     & Chr\tnote{A} & Location  & A1\tnote{B}  & A2\tnote{C} & MAF\tnote{D}   & Mapped genes                       & Sign\tnote{E} & Known\tnote{F} \\ \hline
57  & rs11630918  & 15 & 75155896  & \textbf{C}  & T & 0.383 & SCAMP2                        & $-$ &     \\
58  & rs7187359   & 16 & 30703155  & G  & \textbf{A} & 0.335 & (intergenic variant)          & $+$ &     \\
59  & rs2183405   & 9  & 16661933  & G  & \textbf{A} & 0.271 & BNC2                          & $+$ &     \\
60  & rs2651888   & 1  & 3143384   & G  & \textbf{T} & 0.411 & PRDM16                        & $+$ &     \\
61  & rs2189965   & 7  & 28172014  & \textbf{T}  & C & 0.340 & JAZF1                         & $+$ & Yes \\
62  & rs12911254  & 15 & 75166335  & A  & \textbf{G} & 0.344 & SCAMP2                        & $-$ &     \\
63  & rs757729    & 7  & 28146305  & \textbf{G}  & C & 0.441 & JAZF1                         & $-$ & Yes \\
64  & rs6495122   & 15 & 75125645  & C  & \textbf{A} & 0.478 & CPLX3, ULK3                   & $-$ &     \\
65  & rs4944044   & 11 & 71120213  & A  & \textbf{G} & 0.426 & AP002387.1                    & $-$ &     \\
66  & rs6856032   & 4  & 38763994  & G  & \textbf{C} & 0.109 & RNA5SP158                     & $+$ &     \\
67  & rs1375132   & 2  & 135954405 & \textbf{G}  & A & 0.478 & ZRANB3                        & $+$ &     \\
68  & rs2451138   & 8  & 119238473 & T  & \textbf{C} & 0.314 & SAMD12                        & $-$ &     \\
69  & rs6430538   & 2  & 135539967 & \textbf{T}  & C & 0.470 & CCNT2-AS1                     & $+$ &     \\
70  & rs7651090   & 3  & 185513392 & \textbf{G}  & A & 0.281 & IGF2BP2                       & $+$ & Yes \\
71  & rs4918711   & 10 & 113850019 & T  & \textbf{C} & 0.285 & (intergenic variant)          & $-$ &     \\
72  & rs3861922   & 1  & 198210570 & A  & \textbf{G} & 0.466 & NEK7                          & $-$ &     \\
73  & rs7917983   & 10 & 114732882 & T  & \textbf{C} & 0.481 & TCF7L2                        & $+$ & Yes \\
74  & rs1781145   & 1  & 1388289   & A  & \textbf{C} & 0.362 & ATAD3C                        & $+$ &     \\
75  & rs7170174   & 15 & 94090333  & \textbf{T}  & C & 0.246 & AC091078.1                    & $-$ &     \\
76  & rs7164916   & 15 & 91561446  & T  & \textbf{C} & 0.246 & VPS33B, VPS33B-DT             & $+$ &     \\
77  & rs696859    & 1  & 234656596 & T  & \textbf{C} & 0.430 & (intergenic variant)          & $+$ &     \\
78  & rs28052     & 5  & 142279870 & C  & \textbf{G} & 0.166 & ARHGAP26                      & $+$ &     \\
79  & rs1408799   & 9  & 12672097  & \textbf{T}  & C & 0.277 & (intergenic variant)          & $-$ &     \\
80  & rs10941112  & 5  & 34004707  & \textbf{C}  & T & 0.355 & AMACR, C1QTNF3-AMACR          & $-$ &     \\
81  & rs11856835  & 15 & 74716174  & \textbf{G}  & A & 0.261 & SEMA7A                        & $-$ &     \\
82  & rs4768617   & 12 & 45850022  & \textbf{T}  & C & 0.259 & (intergenic variant)          & $-$ &     \\
83  & rs8012970   & 14 & 101168491 & \textbf{T}  & C & 0.179 & (intergenic variant)          & $-$ &     \\
84  & rs4402960   & 3  & 185511687 & G  & \textbf{T} & 0.187 & IGF2BP2                       & $+$ & Yes \\
85  & rs1695824   & 1  & 1365570   & A  & \textbf{C} & 0.164 & LINC01770, VWA1               & $+$ &     \\
86  & rs934886    & 15 & 55939959  & \textbf{A}  & G & 0.360 & PRTG                          & $-$ &     \\
87  & rs7083429   & 10 & 69303421  & \textbf{G}  & T & 0.367 & CTNNA3                        & $+$ &     \\
88  & rs4918788   & 10 & 114820961 & G  & \textbf{A} & 0.348 & TCF7L2                        & $+$ & Yes \\
89  & rs7219320   & 17 & 17880877  & A  & \textbf{G} & 0.318 & DRC3, AC087163.1, ATPAF2      & $+$ &     \\
90  & rs61822626  & 1  & 205118441 & C  & \textbf{T} & 0.478 & DSTYK                         & $-$ & Yes \\
91  & rs250414    & 5  & 33990623  & \textbf{C}  & T & 0.361 & AMACR, C1QTNF3-AMACR          & $-$ &     \\
92  & rs11073964  & 15 & 91543761  & C  & \textbf{T} & 0.362 & VPS33B,PRC1                   & $+$ & Yes \\
93  & rs17729876  & 10 & 101999746 & \textbf{G}  & A & 0.352 & CWF19L1, SNORA12              & $-$ &     \\
94  & rs2386584   & 15 & 91539572  & T  & \textbf{G} & 0.360 & VPS33B, PRC1                  & $+$ & Yes \\
95  & rs683       & 9  & 12709305  & \textbf{C}  & A & 0.430 & TYRP1, LURAP1L-AS1            & $-$ &     \\
96  & rs17344537  & 1  & 205091427 & \textbf{T}  & G & 0.462 & RBBP5                         & $-$ &     \\
97  & rs10416717  & 19 & 13521528  & A  & \textbf{G} & 0.470 & CACNA1A                       & $+$ &     \\
98  & rs2644590   & 1  & 156875107 & \textbf{C}  & A & 0.453 & PEAR1                         & $-$ &     \\
99  & rs447923    & 5  & 142252257 & \textbf{T}  & C & 0.384 & ARHGAP26, ARHGAP26-AS1        & $+$ &     \\
100 & rs2842895   & 6  & 7106316   & C  & \textbf{G} & 0.331 & RREB1                         & $-$ & Yes \\
101 & rs231354    & 11 & 2706351   & \textbf{C}  & T & 0.329 & KCNQ1, KCNQ1OT1               & $+$ & Yes \\
102 & rs4959424   & 6  & 7084857   & T  & \textbf{G} & 0.410 & (intergenic variant)          & $-$ &     \\
103 & rs2153271   & 9  & 16864521  & T  & \textbf{C} & 0.411 & BNC2                          & $-$ &     \\
104 & rs12142199  & 1  & 1249187   & A  & \textbf{G} & 0.398 & INTS11, PUSL1, ACAP3, MIR6727 & $-$ &     \\
105 & rs2733833   & 9  & 12705095  & \textbf{T}  & G & 0.272 & TYRP1, LURAP1L-AS1            & $-$ &     \\
106 & rs1564782   & 15 & 74622678  & A  & \textbf{G} & 0.283 & CCDC33                        & $-$ &     \\
107 & rs9268644   & 6  & 32408044  & \textbf{C}  & A & 0.282 & HLA-DRA                       & $+$ &     \\
108 & rs271738    & 1  & 234662890 & A  & \textbf{G} & 0.395 & LINC01354                     & $+$ &     \\
109 & rs12907898  & 15 & 75207872  & T  & \textbf{C} & 0.391 & COX5A                         & $-$ &     \\
110 & rs146900823 & 3  & 149192851 & GC & \textbf{G} & 0.344 & TM4SF4                        & $-$ &     \\
111 & rs1635166   & 15 & 28539834  & T  & \textbf{C} & 0.118 & HERC2                         & $-$ &      \\ \hline
\end{tabular}
\begin{tablenotes}
\item[A] Chromosome,
\item[B] Minor allele,
\item[C] Major allele,
\item[D] Minor allele frequency,
\item[E] Sign of the regression coefficient,
\item[F] Mapped gene included in \citet{mahajan2018fine}.
The boldface indicates the risk allele determined by the reference allele and the sign of the regression coefficient. 
\end{tablenotes}
\end{threeparttable}
}
\end{table}

\end{document}